\documentclass[usenatbib]{mn2e}
\usepackage{natbib}
\usepackage{hyperref}
\usepackage{psfig}
\usepackage{epsfig}
\usepackage{pdflscape}
\newif\ifAMStwofonts

\def\sqiglt{\hbox{\rlap{\lower.55ex \hbox {$\sim$}}\kern-.05em \raise.4ex \hbox{$<$}\,}}
\def\sqiggt{\hbox{\rlap{\lower.55ex \hbox {$\sim$}}\kern-.05em \raise.4ex \hbox{$>$}\,}}
\def\til{\ensuremath{\sim\,}}

\def\chisq{\ensuremath{\chi^2}}

\newcommand{\tim}[1]{\ensuremath{\times 10^{#1}}}
\def\deg{\ensuremath{^{\circ}}}

\def\xmm{\emph{XMM}}

\def\cms{\ensuremath{$cm$^{-2}}}

\def\swift{\emph{Swift}}
\def\rosat{\emph{Rosat}}
\def\nh{\ensuremath{N_{\rm H}}}

\def\t0{\ensuremath{T_{0}}}

\def\ic{IceCube}

\def\nu{\ensuremath{\nu}}
\def\ergcms{\ensuremath{$erg cm $^{-2}$ s$^{-1}}}

\title[Swift follow-up of IceCube triggers]{Swift follow-up of IceCube triggers, and implications for the Advanced-LIGO era}

\author[Evans et al.]{P.A. Evans$^1$\thanks{pae9@leicester.ac.uk}, J.P. Osborne$^1$, J.A. Kennea$^2$, M. Smith$^3$, D.M. Palmer$^4$,
\and N. Gehrels$^5$, J.M. Gelbord$^{6,7}$, A. Homeier$^8$, M. Voge$^8$, N.L. Strotjohann$^{8,9}$, 
\and D.F. Cowen$^{10}$, S. B\"oser$^{11}$, M. Kowalski$^{8,9}$, A. Stasik$^9$
\\
$^1$Department of Physics and Astronomy, University of Leicester, Leicester, LE1 7RH, UK
\\
$^2$Department of Astronomy and Astrophysics, Pennsylvania State University,
525 Davey Lab, University Park, PA 16802, USA
\\
$^3$Jet Propulsion Laboratory, 4800 Oak Grove Drive,
Pasadena, CA 91109, USA
\\
$^4$Los Alamos National Laboratory, B244, Los Alamos, NM 87545, USA
\\
$^5$NASA Goddard Space Flight Center, Greenbelt, MD 20771, USA
\\
$^6$Spectral Sciences Inc., 4 Fourth Ave., Burlington, MA 01803, USA
\\
$^7$Eureka Scientific Inc., 2452 Delmer St. Suite 100, Oakland, CA 94602, USA
\\
$^8$Physikalisches Institut, Universit\"at Bonn, Nu{\ss}allee 12, D-53115 Bonn, Germany
\\
$^9$DESY, D-15735 Zeuthen, Germany
\\
$^{10}$Dept. of Astronomy and Astrophysics, Pennsylvania State University, University Park, PA 16802, USA
\\
$^{11}$Institute of Physics, University of Mainz, Staudinger Weg 7, D-55099 Mainz, Germany
\\
}

\date{Accepted 2015 January 17. Received 2015 January 12; in original form 2014 October 24}

\pagerange{\pageref{firstpage}--\pageref{lastpage}}
\pubyear{2015}

\begin{document}

\maketitle

\label{firstpage}

\begin{abstract} 
Between 2011 March and 2014 August  \swift\ responded to 20 triggers from the \ic\ neutrino observatory,
observing the IceCube $50$\%\ confidence error circle in X-rays, typically within 5 hours of the trigger.
No confirmed counterpart has been detected. We describe the \swift\ follow up
strategy and data analysis and present the results of the campaign.
We discuss the challenges of distinguishing the X-ray counterpart to a neutrino 
trigger from serendipitous uncatalogued X-ray sources in the error circle, and consider the implications
of our results for future strategies for multi-messenger astronomy, with particular reference
to the follow up of gravitational wave triggers from the advanced-era detectors.
\end{abstract}


\section{Introduction}
\label{sec:intro}


For biological reasons, astronomy has  been a science carried out using electromagnetic (EM) radiation, and
indeed until comparatively recently was limited to that portion of the
EM spectrum to which our eyes are sensitive, and the atmosphere
transparent. This has changed over the last century and at the present
time observatories exist collecting data from the longest wavelengths
(e.g.\ LOFAR\footnote{http://www.lofar.org}) to the shortest (e.g.\ HESS,
\citealt{Hinton04}). Today, a growing area of astronomical research does
not use EM radation at all, but probes other messengers, such as
neutrinos, gravitational waves or cosmic rays. The detection and identification of
astrophysical sources of these messengers is difficult, and an ideal
scenario is to combine detections of non-EM messengers with EM signals,
to provide a \emph{multi-messenger} dataset. To date, the only object
outside of our solar system to be detected in this way is the supernova
SN1987a \citep{Kunkel87}, which, in addition to its EM discovery and
observations, was also detected as a neutrino emitter
\citep{Alekseev87,Bionta87,Hirata87,Hirata88}. While extra-solar cosmic rays \citep[e.g.][]{Abraham10} and PeV
neutrinos \citep{Aartsen13} have been detected, the origin of these is still unclear and
they have not been reliably coupled with a known EM object.

Neutrinos are expected from various sources, such as supernovae (SNe, 
e.g.\ \citealt{Kachelreiss05,Abbasi14}) and gamma-ray bursts (GRBs, e.g. 
\citealt{Asano14}), but targeted searches -- retrospectively searching 
the neutrino data corresponding to the time and location of EM 
detections of these phenomena -- have so far yielded null results from 
both the IceCube \citep{Abbasi11,Abbasi12} and ANTARES 
\citep{Adrian-Martinez13} observatories. Similarly, gravitational waves 
are expected to arise from a range of phenomena, particularly the merger 
of two neutron stars in a short GRB. Targeted searches for gravitational 
waves from short GRBs have also, so far, failed to produce any 
detections (e.g.\ \citealt{Abadie10b}).

Effort has also been expended to search for EM counterparts to non-EM 
triggers. Because EM facilities tend to have narrow fields of view, the 
likelihood of a non-EM trigger being contemporaneously observed by an EM 
telescope are very low, therefore the EM data have to be collected after 
the non-EM trigger. The error regions from neutrino or gravitational 
wave facilities are on the scale of degrees, thus it often requires 
multiple pointings to collect the necessary EM data. It is also not 
clear when is the optimal time to search for the counterpart, as the 
relative timescales of EM and non-EM radiation depends on the physical 
source of the emission. For example, for supernovae the neutrino signal 
precedes the EM signal by many days. An optimal follow-up facility 
would, therefore, have a large (ideally all-sky) field of view, and high 
level of sensitivity. Due to the high rate of transient events in the 
universe, multiwavelength capabilities are also desirable, for example 
to help distinguish rapidly between GRBs and flare stars.

The \swift\ satellite \citep{GehrelsSwift} arguably provides the best 
existing facility for the EM follow up of non-EM triggers, at least in 
X-rays. Although the X-ray telescope (XRT; \citealt{BurrowsXRT}) has 
only a modest field of view (radius \til0.2\deg), the \swift\ spacecraft 
is capable of rapid slewing, and has the ability to `tile' regions on 
the sky, so as to cover a large error region in a single spacecraft 
orbit. The XRT is sensitive to 5\tim{-13} \ergcms\ in 1 ks (0.3--10 
keV), and can localise sources to a 90\%\ confidence radius of 3.5 
arcsec (improving to 1.4 arcsec for brighter sources; 
\citealt{Goad07,Evans09}). 

\cite{Evans12} reported on \swift\ follow up of two gravitational wave
triggers from the LIGO-Virgo \citep{Abbot09,Accadia12} facilities. No
X-ray counterpart to the gravitational triggers could be found, and
indeed it transpired that neither of the gravitational wave triggers
was in fact real (one was a subthreshold noise event, the other an
artificial signal introduced to the data as a blind test of the
detection algorithms). In this work, we report on the search with
\swift-XRT for X-ray counterparts to 20 neutrino-doublet triggers from
the IceCube facility \citep{Achterberg06}, and discuss the challenges related to idenfitying
the EM counterpart.  A neutrino doublet (or multiplet) was defined as two or more neutrinos detected
within 100 s of each other, and with an angular separation of at most 3.5\deg; more details
about this is given in a companion paper (Aartsen et al., in preparation).

The \swift\ follow-up observations began as soon as possible after the neutrino trigger,
implicitly assuming that the X-ray emission from the astrophysical
neutrino source is temporally coincident with (or only a few hours after) the
neutrino emission. We consider two ways of identifying the X-ray
counterpart: by its brightness compared to reference catalogues, or by
its temporal variability (in particular, whether it shows signs of
fading, as may be expected following some form of outburst).

We did not set the 
threshold at which \swift\ will respond to a neutrino trigger based on theoretical predictions
of neutrino flux (which are highly uncertain due to the lack of observational constraint),
instead we set it such that IceCube would be expected to produce roughly six
spurious (i.e.\ non-astrophysical) triggers per year, which represents a compromise
between sensitivity to astrophysical neutrinos, and the value of \swift's observing time.
The companion paper (Aartsen et al., in preparation) will discuss the expected rate of 
doublet triggers from the background and from astrophysical objects, and consider the 
lack of neutrino triplets during this experiment.

This paper is organised as follows. In Section 2 we describe the follow-up observing strategy employed by 
\swift, and in Section 3 we overview the data analysis techniques. In Section 4 we consider the sources detected,
and attempt to identify if either of these is likely to be the counterpart to the neutrino trigger, which we expect to be 
a source undergoing some form of outburst. Finally, in Section 5 we consider the implications of our findings for future EM follow-up
of non-EM triggers, in particular, the expected gravitational wave triggers from the Advanced LIGO-VIRGO facility.

Throughout the paper we have assumed a cosmology  with $H_0=71$ km s$^{-1}$ Mpc$^{-1}$,
$\Omega_m=0.27, \Omega_{\rm vac}=0.73$. Unless otherwise stated, all quoted errors are at the 90\%\ confidence level,
and upper limits at the 3-$\sigma$ (=99.7\%) confidence level.

\section{\emph{Swift}'s observing strategy}
\label{sec:obs}

\begin{figure}
\begin{center}
\psfig{file=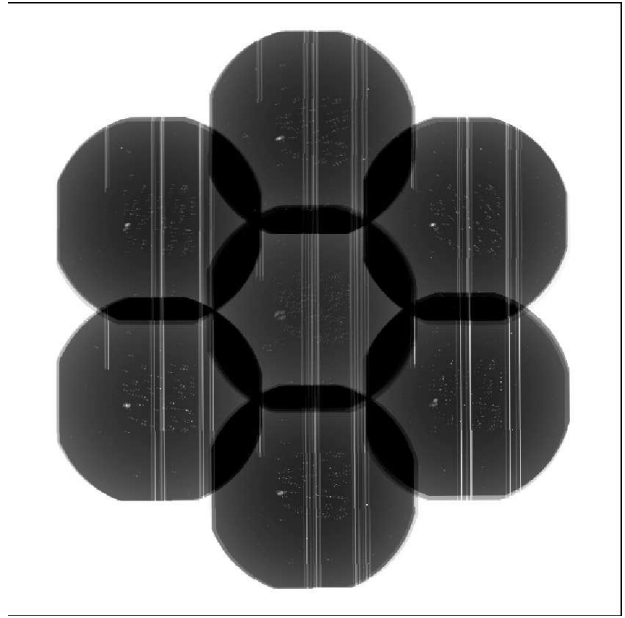,width=8.1cm,angle=-90}
\end{center}
\caption{An example exposure map of a 7-tile \swift-XRT observation of an \ic\ trigger. This observation
was taken with the on-board tiling, so the exposure in each field has been built up over multiple
spacecraft orbits; the pointing is slightly different on each orbit, hence the blurring round the edges 
of the fields. The black lines and dots are the bad columns and pixels on the CCD.}
\label{fig:egexpo}
\end{figure}

\begin{table*}
\caption{Details of the 20 \ic\ triggers followed-up by \swift\ as of 2014 September 03}
\begin{tabular}{ccccc}
\hline
Trigger \#  & Error radius & Trigger time & Delay $^1$ & Tiling type \\
            & (50\% conf) &  (UT) & (hours) \\
\hline
1   & 0.7\deg     &   2011-03-26 21:53:41 & 1.4 & Manual \\
2   & 0.7\deg    &   2011-09-27 12:23:29 & 7.5 & Manual \\
3   & 0.3\deg    &   2011-10-24 02:41:11 & 10.0 & Automatic \\
4   & 0.7\deg    &   2011-12-06 01:40:15 & 1.7 & Automatic \\
5   & 1.1\deg    &   2012-01-17 22:01:34 & 4.8 & Automatic \\
6   & 0.5\deg     &   2012-02-08 00:14:29 & 1.4 & Automatic \\
7   & 1.8\deg      &   2012-03-03 16:47:22 & 1.2 & Automatic \\
8   & 0.7\deg      &   2012-08-29 22:49:59 & 2.5 & Automatic \\
9   & 0.9\deg      &   2012-09-17 18:08:03 & 1.6 & Automatic \\
10  & 0.8\deg     &   2012-10-23 04:46:15 & 5.7 & Automatic \\
11  & 0.4\deg     &   2012-12-21 02:17:24 & 3.7 & Automatic \\
12  & 1.4\deg    &   2013-01-15 11:08:46 & 1.5 & Automatic \\
13  & 0.7\deg    &   2013-02-13 18:47:42 & 1.9 & Automatic \\
14  & 1.2\deg     &   2013-03-08 22:15:49 & 17.0 & Automatic \\
15  & 0.9\deg     &   2013-03-27 19:54:12 & 1.5 & Automatic \\
16  & 1.3\deg    &   2013-05-17 15:57:03 & 0.7 & Automatic \\
17  & 0.8\deg    &   2014-01-08 05:29:08 & 9.0 & Automatic \\
18  & 0.7\deg    &   2014-01-17 04:02:10 & 1.8 & Automatic \\
19  & 0.3\deg    &   2014-02-26 19:04:14 & 3.2 & Automatic \\
20  & 0.7\deg     &   2014-08-29 13:54:49 & 1.1 & Automatic \\
\hline
\end{tabular}
\medskip
\newline $^1$ The time between the possible neutrino event detected by \ic\ and the start
of the first observation with the \swift-XRT.
\label{tab:icobs}
\end{table*}

Following \ic\ triggers, high-priority Target of Opportunity (ToO) requests were submitted
to \swift. Due to the efficient and flexible operation of \swift, observations were able to begin rapidly once the ToO was received:
the median time from \ic\ trigger to the first \swift\ observation was 1.8 hours.
The \ic\ 50\%\ error radius is typically $>0.5\deg$, whereas the \swift-XRT has a field of view of radius of 0.2\deg, therefore
it was necessary to observe the error region in a series of seven overlapping `tiles': an example exposure map is shown
in Fig.~\ref{fig:egexpo}. Initially this tiling had to be
performed by manually commanding seven separate observations as \swift\ Automatic Targets\footnote{That is, the 
observations were not in the pre-planned science timeline, and overrode targets which were. The times
of the observations were set automatically by the on-board software.}; each tile was consequently observed on a separate 
spacecraft orbit. Under this system, all of the requested exposure in a given tile (typically 1--2 ks) was gathered in a single
spacecraft pointing\footnote{XRT can observe a single target for a maximum of 2.7 ks per 96-min spacecraft orbit.};
however, for each successive field the delay between the trigger and the observation increased
by \til96 min (\swift's orbital period). On 2011 August 10 the software on-board \swift\ was modified to support automatic
tiling. In this system, which was used from trigger \#3 onwards (Table~\ref{tab:icobs}), a single Automatic Target is uploaded, 
and \swift\ automatically divides the visibility window of the IceCube trigger location in each spacecraft orbit between the 7 tiles.
This is repeated on each orbit until the requested exposure time has been gathered for each tile\footnote{The minimum
time permitted for a continuous exposure is 60 s; if the observing window is not sufficient for all tiles to be observed for
at least this duration on a given orbit, then the next orbit will begin with the first unobserved tile.}. Under this system,
all tiles are usually observed in the first visibility window after the observation is uploaded, however the
individual exposures are short, so for a given tile to accumulate its full exposure time takes longer. This 
strategy is better than the manual-upload approach,  since it allows \swift\ to cover the entire search box promptly, increasing the chances of 
detecting a bright, rapidly-fading counterpart (e.g.\ a GRB afterglow), without significantly reducing the likelihood 
of finding a slowly-fading counterpart. Details of the 20 IceCube triggers are given in Table~\ref{tab:icobs}.

\section{Data analysis}
\label{sec:data}

The XRT data were automatically analysed at the United Kingdom Swift 
Science Data Centre (UKSSDC) at the University of Leicester, using {\sc 
heasoft} v6.15.1, and the \swift\ {\sc caldb} 
b20090130\_u20111031\_x20131220\_m20140221. The data were first processed 
using {\sc xrtpipeline}, then a series of source detection and 
analysis routines were applied. In a previous work, similar to this but 
relating to gravitational wave triggers \citep{Evans12}, we noted the need 
for a source detection system that was optimised for fainter sources. Since then 
such a system has been developed \citep{Evans14}, and we used it in this work. 
It consists of the following steps: filtering the data, creating images and exposure
maps\footnote{This step differed in 
one detail from \cite{Evans14}: in that work images could not exceed 
$1000\times1000$ pixels in size. We have 
overcome this limitation so a single image covering all 7 tiles and being 
$1900\times1900$ pixels ($1.25\deg\times1.25\deg$) in size is used.}, and locating and characterising sources.
The latter step is an iterative process which uses a 
combination of sliding-cell detection, background modelling, source 
PSF-fitting and likelihood tests to detect and localise sources. It also 
provides a quality flag for each source, which indicates the probability 
of the source being spurious. 0.3\%\ of sources flagged as \emph{Good} 
are spurious; this rises to 1\%\ when the \emph{Good} and 
\emph{Reasonable} sources are considered (\emph{Reasonable} sources on 
their own have a 7\%\ false positive rate), and 10\%\ when all sources 
(\emph{Good}, \emph{Reasonable} and \emph{Poor}) are included 
(\emph{Poor} sources on their own have a 35\%\ false positive rate, so 
should be viewed with caution). Full details of this procedure are given in 
\cite{Evans14}, particularly sections 3.4 and 7 and fig.~3.

The astrometric accuracy of \swift-XRT positions, determined using the on-board star trackers,
is 3.5 arcsec (90\% confidence; \citealt{Moretti07}). 
This can sometimes be improved either by using the UV/optical telescope on \swift\ as a super star-tracker
(so-called `position enhancement'; \citealt{Goad07,Evans09}), or by aligning detected
XRT sources with 2MASS \citep{Skrutskie06} objects (\citealt{Butler07d,Evans14}). We tried both of these
methods for each position, although in most cases the source was too faint for the former method to work
(i.e.\ the XRT source could not be detected in an exposure corresponding to a single UVOT image), and
there were too few XRT/2MASS matches for the latter method to offer a more accurate astrometric solution
than the star-trackers on \swift.

\begin{figure}
\begin{center}
\psfig{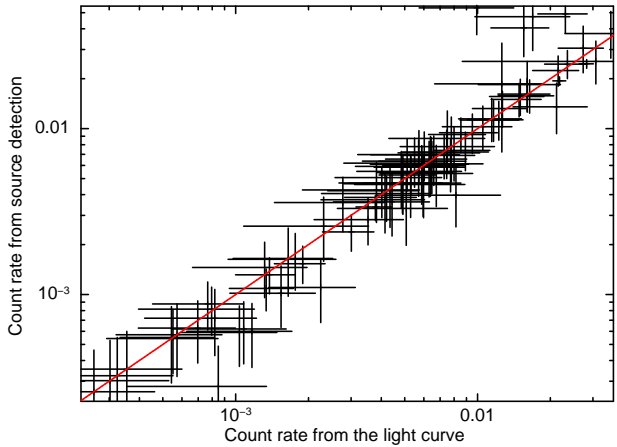}
\end{center}
\caption{The mean source count-rates determined from the source detection plotted against the values derived
from source light curves, where both are available. The two are in good agreement below 0.01 ct s$^{-1}$, confirming
that using the brightness estimates from source detection is a safe approach for the faintest sources for
which light curves could not be produced.}
\label{fig:lccheck}
\end{figure}

We also built light curves and spectra of each source using the tools developed by \cite{Evans07,Evans09}.  These provide
fully validated products in which all necessary corrections (e.g.\ for pile up -- where multiple photons impact a single pixel
in a single CCD frame, and are consequently miscounted as a single event -- the presence of dead columns on the XRT CCD, and exposure
variation across the image) have been applied. For most objects there were too few counts to produce more than a single light-curve bin,
thus the `light curve' is really simply a single brightness measurement. For those objects with sufficient counts, we binned the light curve 
such that there were at least 15 counts in each bin. 
The spectra were automatically fitted in {\sc xspec} \citep{Arnaud96}
with an absorbed power-law model, using two {\sc tbabs} absorption components (with redshift fixed at 0).
One of these was fixed at the Galactic value of \nh\ from \cite{Willingale13}, the other was free to vary.
The spectral fit was used to determine the energy conversion factor (ECF; conversion from the XRT count rate into an observed 0.3--10 keV flux)
for each individual object, and also, via {\sc pimms}, a conversion from the measured XRT count-rate
to the equivalent in the \rosat\ PSPC (which had an energy coverage of 0.1--2.4 keV, \citealt{Pfeffermann87}). The latter conversion allows us 
to compare the detected sources with the \rosat\ all-sky survey (RASS; \citealt{Voges99}): this is the most sensitive all-sky X-ray survey to
date and we use it as a reference catalogue (see Section~\ref{sec:counterpart}),
along with the  \xmm\ Slew Survey (XSS; \citealt{Saxton08,Warwick12}) which 
covers \til2/3 of the sky in the 0.2--10 keV band.

For some sources, a light curve or spectrum could not be produced (or the spectrum could not be fitted) as the source was too faint, or
too near to the edge of the field of view. Where the light curve could not be produced, the mean count-rate was
taken from the output of the source detection procedure (which includes corrections for PSF losses, pile up etc.). These were designed 
as indicators, and are not fully calibrated. Therefore, to verify their accuracy we compared the mean count-rate from the light curve
with that from the source detection routine for all sources where both were available. As Fig.~\ref{fig:lccheck} shows, the two are in reasonable agreement. 

For objects where the automatic spectrum or fit could not be produced, we still require both an ECF and a conversion between the XRT count rate and an equivalent
\rosat\ PSPC rate, to compare the source brightness with the XSS and RASS respectively. In these cases we used a standard AGN spectral model: an absorbed power-law 
with \nh=3\tim{20} \cms, and $\Gamma=1.7$. This yields a 0.3--10 keV ECF of 4.03\tim{-11} erg \cms\ XRT ct$^{-1}$, and
a conversion of 1.497 PSPC counts per XRT count.

In Table~\ref{tab:sources} we list all of the X-ray objects detected in 
our follow-up observations, giving their positions, the mean 0.3--10 keV 
flux, and various parameters which are useful in identifying whether the 
source is related to the neutrino trigger (described in 
Section~\ref{sec:counterpart}). The sources for which a spectrum could 
be constructed and fitted are shown in Table~\ref{tab:srcspec} along 
with details of the spectral fit. A small number of sources have 
unconstrained spectral fit parameters, or extreme values which are 
artefacts of a poor quality spectrum, rather than indicating extreme 
physical properties. However, since the ECF and PSPC conversion for 
those sources are compatible with the range seen for the other sources, 
we do not revert to the standard values given above. One object, source 
2 in the field of trigger 1, has a very high ECF which is driven by the 
fact that the spectrum is very hard, however we have not reverted to the 
standard AGN spectrum for this, as it is clearly inappropriate. The lack 
of soft emission from this object means that we do not expect \rosat\ to 
have seen it; a fact which would be unclear if we used the standardised 
spectrum: that is only used for objects where we had no XRT spectral fit 
at all. 

\begin{landscape}

\begin{table}
\caption{Sources detected in the XRT follow up of the IceCube triggers.}
\label{tab:sources}
\begin{tabular}{cccccccccccc}
\hline
IceCube & Src \# & Location  & Err$^a$ & Flag$^b$ & Exposure &  C (B)$^c$ & Obs Flux$^d$   & Upper Limit$^e$ & $N_{\rm seren}^f$ & $P_{\rm const}^g$ & Cat?$^h$ \\
Field   &           & (J2000)   & (${\prime\prime}$)  &  & (ks) &  & (\tim{-13} \ergcms) & (\tim{-13} \ergcms)      &  & &   \\
        &           &           &           & & & & (0.3--10 keV) & (0.3--10 keV) & & &  \\
\hline
1 & 1 & 03$^h$ 31$^m$ 05.$^s$1 +13$\deg$ 56$^\prime$ 00$^{\prime\prime}$ & 3.2$^*$ & G & 3.5 & 16 (0.8) & 1.81 ($\pm$0.48) & 12 & 2.7 &  & \\
1 & 2 & 03$^h$ 30$^m$ 58.$^s$7 +13$\deg$ 57$^\prime$ 06$^{\prime\prime}$ & 5.2 & G & 3.5 & 12 (0.8) & 1940 ($\pm$640) & $\S$ & 3.7 &  & \\
1 & 3$^\mathparagraph$ & 03$^h$ 31$^m$ 01.$^s$8 +14$\deg$ 15$^\prime$ 06$^{\prime\prime}$ & 6.1$^*$ & G & 1.8 & 8 (0.5) & 1.97 ($\pm$0.84) & 6.9 & 3.1 &  & \\
1 & 4 & 03$^h$ 29$^m$ 47.$^s$9 +14$\deg$ 26$^\prime$ 37$^{\prime\prime}$ & 4.4$^*$ & G & 1.5 & 8 (0.5) & 3.8 ($\pm$1.6) & 33 & 2.1 &  & \\
3 & 1$^\mathparagraph$ & 14$^h$ 59$^m$ 06.$^s$8 +59$\deg$ 31$^\prime$ 09$^{\prime\prime}$ & 4.3$^*$ & G & 1.3 & 17 (0.5) & 5.1 ($\pm$1.3) & & 0.7 &  & Y\\
3 & 2 & 14$^h$ 59$^m$ 05.$^s$1 +60$\deg$ 08$^\prime$ 43$^{\prime\prime}$ & 3.7$^*$ & G & 1.8 & 10 (0.6) & 3.1 ($\pm$1.2) & 18 & 2.3 &  & \\
3 & 3$^\mathparagraph$ & 15$^h$ 01$^m$ 06.$^s$8 +59$\deg$ 59$^\prime$ 43$^{\prime\prime}$ & 5.7 & G & 2.1 & 7 (0.8) & 1.79 ($\pm$0.76) & 7.6 & 2.6 &  & \\
4 & 1$^\mathparagraph$ & 22$^h$ 15$^m$ 02.$^s$5 +57$\deg$ 02$^\prime$ 43$^{\prime\prime}$ & 3.5$^*$ & G & 0.4 & 16 (0.1) & 9.7 ($\pm$2.7); 9.3 ($\pm$2.1) & 15 & 0.1 & 0.78 & \\
4 & 2 & 22$^h$ 15$^m$ 03.$^s$2 +57$\deg$ 02$^\prime$ 16$^{\prime\prime}$ & 4.0$^*$ & G & 0.4 & 10 (0.1) & 4.1 ($\pm$1.4); 4.03 ($\pm$0.85) & 4.8 & 0.1 & 0.92 & \\
5 & 1$^\mathparagraph$ & 08$^h$ 31$^m$ 36.$^s$4 +18$\deg$ 20$^\prime$ 55$^{\prime\prime}$ & 4.3$^*$ & R & 0.2 & 15 (3.5) & 4.0 ($\pm$1.7) & 12 & 0.3 &  & \\
5 & 2$^\mathparagraph$ & 08$^h$ 31$^m$ 10.$^s$8 +18$\deg$ 33$^\prime$ 01$^{\prime\prime}$ & 3.2$^*$ & G & 0.3 & 14 (3.6) & 6.3 ($\pm$1.7) & 13 & 0.3 &  & \\
5 & 3$^\mathparagraph$ & 08$^h$ 31$^m$ 09.$^s$6 +18$\deg$ 25$^\prime$ 37$^{\prime\prime}$ & 3.2$^*$ & G & 0.8 & 11 (4.4) & 3.8 (+2.1, -1.7)$^\sharp$ & 10 & 0.3 &  & \\
5 & 4$^\mathparagraph$ & 08$^h$ 32$^m$ 12.$^s$8 +18$\deg$ 37$^\prime$ 02$^{\prime\prime}$ & 3.2$^*$ & G & 0.7 & 15 (3.8) & 6.5 ($\pm$2.4) & 14 & 0.3 &  & \\
5 & 5$^\mathparagraph$ & 08$^h$ 31$^m$ 38.$^s$0 +18$\deg$ 37$^\prime$ 55$^{\prime\prime}$ & 3.2$^*$ & R & 0.2 & 7 (2.7) & 8.0 (+5.6, -4.3)$^\sharp$ & 17 & 0.2 &  & \\
5 & 6$^\mathparagraph$ & 08$^h$ 30$^m$ 47.$^s$0 +18$\deg$ 57$^\prime$ 09$^{\prime\prime}$ & 5.7 & G & 0.6 & 6 (0.2) & 4.8 (+2.3, -1.8)$^\sharp$ & 18 & 0.4 &  & Y\\
5 & 7 & 08$^h$ 30$^m$ 28.$^s$9 +18$\deg$ 46$^\prime$ 11$^{\prime\prime}$ & 5.8 & R & 0.4 & 11 (3.0) & 6.6 (+3.1, -2.5)$^\sharp$ & 5.4 & 0.2 &  & \\
5 & 8$^\mathparagraph$ & 08$^h$ 30$^m$ 28.$^s$8 +18$\deg$ 46$^\prime$ 10$^{\prime\prime}$ & 4.0$^*$ & R & 0.3 & 9 (4.7) & 7.5 (+5.7, -4.6)$^\sharp$ & 6.3 & 0.2 &  & \\
5 & 9$^\mathparagraph$ & 08$^h$ 30$^m$ 04.$^s$3 +18$\deg$ 52$^\prime$ 31$^{\prime\prime}$ & 4.7$^*$ & R & 0.2 & 13 (4.3) &19.1 (+8.6, -7.1)$^\sharp$ & 8.9 & 0.09 &  & \\
6 & 1$^\mathparagraph$ & 01$^h$ 27$^m$ 33.$^s$8 +10$\deg$ 24$^\prime$ 09$^{\prime\prime}$ & 5.0$^*$ & G & 1.4 & 12 (0.5) & 4.2 ($\pm$1.4) & 13 & 1.0 &  & \\
6 & 2 & 01$^h$ 25$^m$ 51.$^s$4 +09$\deg$ 59$^\prime$ 24$^{\prime\prime}$ & 5.5 & G & 1.6 & 10 (0.5) & 2.21 ($\pm$0.82) & 44 & 1.7 &  & \\
6 & 3 & 01$^h$ 24$^m$ 25.$^s$8 +10$\deg$ 09$^\prime$ 32$^{\prime\prime}$ & 5.5 & G & 1.6 & 9 (0.6) & 2.9 ($\pm$1.3) & 29 & 3.0 &  & \\
6 & 4 & 01$^h$ 24$^m$ 45.$^s$2 +10$\deg$ 07$^\prime$ 08$^{\prime\prime}$ & 7.0 & G & 1.6 & 9 (0.4) & 3.2 ($\pm$1.2) & 32 & 2.1 &  & \\
6 & 5$^\mathparagraph$ & 01$^h$ 24$^m$ 51.$^s$5 +10$\deg$ 10$^\prime$ 33$^{\prime\prime}$ & 6.1 & G & 1.7 & 7 (0.4) & 1.98 ($\pm$0.90) & 13 & 3.1 &  & \\
6 & 6$^\mathparagraph$ & 01$^h$ 25$^m$ 23.$^s$1 +10$\deg$ 03$^\prime$ 23$^{\prime\prime}$ & 8.4 & G & 1.5 & 9 (0.5) & 2.3 ($\pm$1.0) &  & 2.5 &  & Y\\
6 & 7 & 01$^h$ 26$^m$ 56.$^s$8 +10$\deg$ 12$^\prime$ 59$^{\prime\prime}$ & 5.7 & G & 2.0 & 7 (0.6) & 1.39 ($\pm$0.63) & 38 & 3.5 &  & \\
6 & 8$^\mathparagraph$ & 01$^h$ 26$^m$ 12.$^s$4 +10$\deg$ 26$^\prime$ 27$^{\prime\prime}$ & 10.4 & G & 1.7 & 7 (0.6)  & 2.4 ($\pm$1.2) & 11 & 2.3 &  & \\
6 & 9$^\mathparagraph$ & 01$^h$ 25$^m$ 52.$^s$7 +10$\deg$ 22$^\prime$ 14$^{\prime\prime}$ & 7.7 & R & 2.8 & 7 (0.9) & 0.93 ($\pm$0.50) & 9.2 & 1.4 &  & \\
6 & 10 & 01$^h$ 24$^m$ 51.$^s$1 +10$\deg$ 33$^\prime$ 06$^{\prime\prime}$ & 4.3$^*$ & G & 1.5 & 21 (0.5) & 4.8 ($\pm$2.4); 4.2 ($\pm$1.0) & 17 & 0.5 & 0.71 & \\
6 & 11 & 01$^h$ 26$^m$ 11.$^s$4 +10$\deg$ 23$^\prime$ 37$^{\prime\prime}$ & 6.2 & G & 2.6 & 9 (0.9) & 0.84 ($\pm$0.40) & 9.0 & 1.3 &  & \\
7 & 1 & 04$^h$ 08$^m$ 41.$^s$0 +03$\deg$ 34$^\prime$ 37$^{\prime\prime}$ & 1.5$^*$ & G & 12.9 & 318 (12.6) & 11.3 ($\pm$2.1); 5.43 ($\pm$0.36) &  & 0.2 & 5.6\tim{-4} & Y\\
\hline
\end{tabular}
\medskip
\newline $^a$ radius, 90\% confidence.
\newline $^b$ G=Good, R=Reasonable, P=Poor; see text for details.
\newline $^c$ Number of counts found in the region used to detect the source, with the number expected from the background in parentheses. 
\newline $^d$ This is the observed (i.e.\ absorbed) flux. For objects with more than one light curve bin, the peak flux and mean flux are given.
\newline $^e$ 3 $\sigma$ upper limit, from the RASS unless otherwise indicated. If no value is given, this is because the source
is a known X-ray emitter, so an upper limit is unnecessary.
\newline $^f$ The number of sources at least as bright as this one, expected in the XRT field of view. See Section~\ref{sec:pseren} for details.
\newline $^g$ The probability that the source is constant, from a \chisq\ test. Only available if the light curve contained at least 2 bins.
\newline $^h$ Whether a catalogued source was found which matched this object; details are given in Table~\ref{tab:catmatch}
\newline $^*$ Position enhanced by UVOT astrometry.
\newline $^\dagger$ Position derived from XRT-2MASS astrometry.
\newline $^\ddagger$ Upper limit obtained from the \xmm\ Slew Survey.
\newline $^\mathparagraph$ No XRT spectral fit was available, generic AGN spectrum assumed.
\newline $^\sharp$ No XRT light curve was available: the brightness was estimated by the source detection routine.
\newline $^\S$ This was a very hard source to which \rosat\ was insensitive, therefore a RASS limit is uninformative.
\end{table}        

\clearpage

\begin{table}
\contcaption{}
\begin{tabular}{cccccccccccc}
\hline
IceCube & Src \# & Location  & Err$^a$ & Flag$^b$ & Exposure &  C (B)$^c$ & Obs Flux$^d$   & Upper Limit$^e$ & $N_{\rm seren}^f$ & $P_{\rm const}^g$ & Cat?$^h$ \\
Field   &           & (J2000)   & (${\prime\prime}$)  &  & (ks) &  & (\tim{-13} \ergcms) & (\tim{-13} \ergcms)      &  & &   \\
        &           &           &           & & & & (0.3--10 keV) & (0.3--10 keV) & & &  \\
\hline

7 & 2 & 04$^h$ 09$^m$ 25.$^s$0 +03$\deg$ 12$^\prime$ 37$^{\prime\prime}$ & 6.2 & G & 2.6 & 10 (0.8) & 1.66 ($\pm$0.67) & 12 & 2.8 &  & \\
7 & 3$^\mathparagraph$ & 04$^h$ 09$^m$ 42.$^s$9 +03$\deg$ 19$^\prime$ 20$^{\prime\prime}$ & 4.8 & G & 2.1 & 6 (0.6) & 2.04 ($\pm$0.98) & 15 & 2.9 &  & Y\\
7 & 4 & 04$^h$ 09$^m$ 27.$^s$0 +03$\deg$ 34$^\prime$ 07$^{\prime\prime}$ & 1.5$^*$ & G & 18.2 & 404 (15.7) & 22.3 ($\pm$5.8); 7.91 ($\pm$0.49) & 12 & 0.3 & 2.6\tim{-2} & \\
7 & 5 & 04$^h$ 10$^m$ 40.$^s$7 +03$\deg$ 19$^\prime$ 19$^{\prime\prime}$ & 5.5 & G & 1.7 & 9 (0.6) & 1.26 ($\pm$0.47) & 4.2 & 2.1 &  & \\
7 & 6 & 04$^h$ 09$^m$ 53.$^s$2 +03$\deg$ 18$^\prime$ 48$^{\prime\prime}$ & 7.7 & G & 2.0 & 7 (0.8) & 1.02 ($\pm$0.49) & 16 & 3.0 &  & \\
7 & 7 & 04$^h$ 08$^m$ 42.$^s$9 +03$\deg$ 39$^\prime$ 01$^{\prime\prime}$ & 4.2 & G & 14.3 & 46 (4.7) & 1.83 ($\pm$0.36); 1.57 ($\pm$0.26) & 35 & 3.0 & 0.10 & \\
7 & 8 & 04$^h$ 09$^m$ 27.$^s$8 +03$\deg$ 30$^\prime$ 16$^{\prime\prime}$ & 4.3 & G & 19.0 & 27 (6.1) & 0.94 ($\pm$0.41); 0.92 ($\pm$0.25) & 112 & 5.1 & 0.95 & \\
7 & 9 & 04$^h$ 09$^m$ 16.$^s$7 +03$\deg$ 40$^\prime$ 19$^{\prime\prime}$ & 5.4 & G & 17.6 & 23 (5.7) & 1.20 ($\pm$0.43); 1.05 ($\pm$0.31) & 44 & 5.8 & 0.46 & \\
7 & 10 & 04$^h$ 09$^m$ 58.$^s$5 +03$\deg$ 27$^\prime$ 42$^{\prime\prime}$ & 5.3 & G & 9.5 & 26 (3.9) & 0.74 ($\pm$0.19); 0.70 ($\pm$0.17) & 7.0 & 2.4 & 0.43 & \\
7 & 11 & 04$^h$ 08$^m$ 29.$^s$6 +03$\deg$ 33$^\prime$ 03$^{\prime\prime}$ & 5.2 & G & 4.8 & 13 (2.0) & 1.32 ($\pm$0.53) & 25 & 2.2 &  & \\
7 & 12 & 04$^h$ 09$^m$ 03.$^s$5 +03$\deg$ 29$^\prime$ 36$^{\prime\prime}$ & 5.9 & G & 17.3 & 16 (6.1) & 0.29 ($\pm$0.13) & 5.6 & 6.3 &  & \\
7 & 13 & 04$^h$ 09$^m$ 21.$^s$0 +03$\deg$ 38$^\prime$ 04$^{\prime\prime}$ & 5.5 & G & 18.3 & 22 (5.8) & 0.80 ($\pm$0.40) & 104 & 1.2 &  & \\
7 & 14 & 04$^h$ 09$^m$ 08.$^s$9 +03$\deg$ 31$^\prime$ 36$^{\prime\prime}$ & 5.3 & G & 19.1 & 17 (6.2) & 0.53 ($\pm$0.23) & 22 & 8.4 &  & \\
7 & 15$^\mathparagraph$ & 04$^h$ 09$^m$ 36.$^s$7 +03$\deg$ 31$^\prime$ 59$^{\prime\prime}$ & 5.2 & G & 17.8 & 14 (5.6) & 0.44 ($\pm$0.22) & 9.9 & 7.5 &  & \\
7 & 16$^\mathparagraph$ & 04$^h$ 09$^m$ 54.$^s$8 +03$\deg$ 33$^\prime$ 47$^{\prime\prime}$ & 7.0 & P & 16.6 & 16 (5.7) & 0.30 ($\pm$0.16); 0.28 ($\pm$0.12) & 6.2 & 13.6 & 0.69 & \\
7 & 17 & 04$^h$ 08$^m$ 43.$^s$0 +03$\deg$ 33$^\prime$ 50$^{\prime\prime}$ & 5.9 & R & 14.8 & 23 (5.1) & 1.59 ($\pm$0.42); 1.32 ($\pm$0.31) & 175 & 3.5 & 0.12 & \\
7 & 18 & 04$^h$ 09$^m$ 12.$^s$3 +03$\deg$ 42$^\prime$ 34$^{\prime\prime}$ & 5.0 & G & 17.2 & 19 (5.8) & 0.33 ($\pm$0.18); 0.33 ($\pm$0.13) & 14 & 6.8 & 0.98 & \\
7 & 19 & 04$^h$ 09$^m$ 40.$^s$6 +03$\deg$ 46$^\prime$ 28$^{\prime\prime}$ & 7.0 & P & 7.3 & 11 (2.3) & 0.42 ($\pm$0.21) & 9.6 & 5.6 &  & \\
7 & 20 & 04$^h$ 09$^m$ 24.$^s$9 +03$\deg$ 46$^\prime$ 25$^{\prime\prime}$ & 10.8 & G & 9.8 & 11 (3.2) & 0.75 ($\pm$0.45); 0.61 ($\pm$0.24) & 12 & 4.7 & 0.58 & \\
7 & 21$^\mathparagraph$ & 04$^h$ 09$^m$ 59.$^s$2 +03$\deg$ 44$^\prime$ 11$^{\prime\prime}$ & 9.7 & R & 6.2 & 12 (2.1) & 0.71 ($\pm$0.34) & 11 & 5.6 &  & \\
7 & 22 & 04$^h$ 09$^m$ 16.$^s$7 +03$\deg$ 29$^\prime$ 38$^{\prime\prime}$ & 5.3 & G & 18.4 & 15 (6.1) & 0.54 (+0.26, -0.21)$^\sharp$ & 39 & 17.5 &  & \\
7 & 23 & 04$^h$ 10$^m$ 10.$^s$3 +03$\deg$ 33$^\prime$ 34$^{\prime\prime}$ & 6.8 & R & 8.0 & 10 (3.0) & 0.65 ($\pm$0.39); 0.39 ($\pm$0.15) & 6.5 & 2.8 & 0.12 & \\
7 & 24$^\mathparagraph$ & 04$^h$ 08$^m$ 28.$^s$1 +03$\deg$ 37$^\prime$ 55$^{\prime\prime}$ & 8.8 & R & 4.1 & 7 (1.2) & 0.66 (+0.35, -0.27)$^\sharp$ & 13 & 4.0 &  & \\
7 & 25$^\mathparagraph$ & 04$^h$ 09$^m$ 43.$^s$6 +03$\deg$ 28$^\prime$ 35$^{\prime\prime}$ & 5.9 & P & 17.7 & 10 (6.2) & 0.122 (+0.091, -0.074)$^\sharp$ & 7.9 & 16.6 &  & \\
7 & 26$^\mathparagraph$ & 04$^h$ 10$^m$ 00.$^s$1 +03$\deg$ 36$^\prime$ 23$^{\prime\prime}$ & 6.2 & P & 15.1 & 6 (1.4) & 0.23 (+0.12, -0.10)$^\sharp$ & 7.8 & 13.9 &  & \\
7 & 27$^\mathparagraph$ & 04$^h$ 09$^m$ 30.$^s$9 +03$\deg$ 29$^\prime$ 53$^{\prime\prime}$ & 5.4 & R & 19.0 & 12 (6.1) & 0.34 ($\pm$0.20) & 8.7 & 9.9 &  & \\
7 & 28$^\mathparagraph$ & 04$^h$ 09$^m$ 21.$^s$7 +03$\deg$ 27$^\prime$ 40$^{\prime\prime}$ & 6.5 & P & 17.0 & 12 (5.7) & 0.143 (+0.099, -0.080)$^\sharp$ & 12 & 22.0 &  & \\
7 & 29$^\mathparagraph$ & 04$^h$ 09$^m$ 16.$^s$1 +03$\deg$ 39$^\prime$ 06$^{\prime\prime}$ & 5.1 & R & 18.1 & 11 (5.7) & 0.131 (+0.091, -0.074)$^\sharp$ & 7.1 & 20.1 &  & \\
7 & 30$^\mathparagraph$ & 04$^h$ 08$^m$ 54.$^s$9 +03$\deg$ 27$^\prime$ 31$^{\prime\prime}$ & 8.1 & P & 15.1 & 6 (1.1) & 0.139 (+0.078, -0.064)$^\sharp$ & 10 & 18.5 &  & \\
7 & 31$^\mathparagraph$ & 04$^h$ 10$^m$ 01.$^s$5 +03$\deg$ 30$^\prime$ 24$^{\prime\prime}$ & 7.1 & P & 13.0 & 13 (4.7) & 0.52 ($\pm$0.39); 0.33 ($\pm$0.16) & 10 & 12.5 & 0.31 & \\
7 & 32$^\mathparagraph$ & 04$^h$ 09$^m$ 06.$^s$0 +03$\deg$ 37$^\prime$ 12$^{\prime\prime}$ & 5.3 & P & 19.8 & 6 (2.1) & 0.105 (+0.082, -0.067)$^\sharp$ & 8.9 & 2.86 &  & \\
9 & 1 & 04$^h$ 03$^m$ 51.$^s$5 +05$\deg$ 10$^\prime$ 47$^{\prime\prime}$ & 4.7$^{\dagger}$ & G & 1.8 & 10 (0.2) & 3.19 ($\pm$0.86) & 16 & 1.1 &  & \\
9 & 2$^\mathparagraph$ & 04$^h$ 03$^m$ 06.$^s$3 +04$\deg$ 44$^\prime$ 13$^{\prime\prime}$ & 7.1 & G & 0.5 & 8 (0.1)  & 8.4 ($\pm$3.4) &  & 0.3 &  & Y\\
9 & 3 & 04$^h$ 03$^m$ 43.$^s$1 +05$\deg$ 16$^\prime$ 28$^{\prime\prime}$ & 5.9$^{\dagger}$ & G & 1.6 & 8 (0.7) & 0.95 ($\pm$0.44) & 10 & 3.2 &  & \\
9 & 4$^\mathparagraph$ & 04$^h$ 04$^m$ 28.$^s$5 +05$\deg$ 18$^\prime$ 44$^{\prime\prime}$ & 4.3$^{\dagger}$ & G & 1.5 & 15 (0.5) & 4.9 ($\pm$1.3) &  & 0.8 &  & Y\\
9 & 5$^\mathparagraph$ & 04$^h$ 03$^m$ 09.$^s$6 +05$\deg$ 27$^\prime$ 27$^{\prime\prime}$ & 5.9$^{\dagger}$ & G & 1.5 & 6 (0.3) & 1.70 ($\pm$0.85) & 10 & 3.4 &  & \\
10 & 1$^\mathparagraph$ & 22$^h$ 39$^m$ 49.$^s$1 +06$\deg$ 06$^\prime$ 00$^{\prime\prime}$ & 5.5 & G & 0.3 & 10 (0.2) & 14.3 ($\pm$5.1) &  & 0.1 &  & Y\\
10 & 2$^\mathparagraph$ & 22$^h$ 40$^m$ 32.$^s$6 +06$\deg$ 39$^\prime$ 59$^{\prime\prime}$ & 5.0 & G & 1.3 & 13 (0.3) & 4.6 (+1.5, -1.3)$^\sharp$ & 18 & 0.8 &  & \\
10 & 3$^\mathparagraph$ & 22$^h$ 38$^m$ 49.$^s$7 +06$\deg$ 31$^\prime$ 10$^{\prime\prime}$ & 6.5 & G & 0.9 & 6 (0.4) & 3.1 (+1.6, -1.2)$^\sharp$ & 13 & 1.0 &  & \\
11 & 1$^\mathparagraph$ & 14$^h$ 14$^m$ 58.$^s$2 +07$\deg$ 41$^\prime$ 49$^{\prime\prime}$ & 5.8 & G & 1.7 & 10 (0.7) & 2.66 ($\pm$0.98) & 7.0 & 1.1 &  & \\
\hline
\end{tabular}
\end{table}

\clearpage

\begin{table}
\contcaption{}
\begin{tabular}{cccccccccccc}
\hline
IceCube & Src \# & Location  & Err$^a$ & Flag$^b$ & Exposure &  C (B)$^c$ & Obs Flux$^d$   & Upper Limit$^e$ & $N_{\rm seren}^f$ & $P_{\rm const}^g$ & Cat?$^h$ \\
Field   &           & (J2000)   & (${\prime\prime}$)  &  & (ks) &  & (\tim{-13} \ergcms) & (\tim{-13} \ergcms)      &  & &   \\
        &           &           &           & & & & (0.3--10 keV) & (0.3--10 keV) & & &  \\
\hline

11 & 2$^\mathparagraph$ & 14$^h$ 15$^m$ 47.$^s$1 +08$\deg$ 08$^\prime$ 10$^{\prime\prime}$ & 3.3$^*$ & G & 1.3 & 28 (0.8) & 9.5 ($\pm$1.8) &  & 0.1 &  & Y\\
12 & 1$^\mathparagraph$ & 23$^h$ 47$^m$ 47.$^s$9 +20$\deg$ 32$^\prime$ 02$^{\prime\prime}$ & 7.6 & G & 0.4 & 6 (0.1) & 7.1 (+3.4, -2.6)$^\sharp$ & 5.2 & 0.4 &  & \\
13 & 1 & 10$^h$ 57$^m$ 46.$^s$9 +36$\deg$ 15$^\prime$ 39$^{\prime\prime}$ & 6.1 & G & 0.8 & 9 (0.3) & 8.3 ($\pm$2.8) & 16 & 0.3 &  & Y\\
13 & 2$^\mathparagraph$ & 10$^h$ 58$^m$ 17.$^s$1 +35$\deg$ 30$^\prime$ 30$^{\prime\prime}$ & 6.8 & G & 1.7 & 9 (0.6) & 2.40 (+0.96, -0.77)$^\sharp$ & 19 & 2.3 &  & \\
13 & 3$^\mathparagraph$ & 10$^h$ 57$^m$ 02.$^s$8 +35$\deg$ 35$^\prime$ 22$^{\prime\prime}$ & 8.6 & G & 2.5 & 10 (1.0) & 1.72 (+0.68, -0.55)$^\sharp$ & 10 & 3.4 &  & Y\\
13 & 4$^\mathparagraph$ & 10$^h$ 56$^m$ 31.$^s$9 +35$\deg$ 41$^\prime$ 52$^{\prime\prime}$ & 5.8 & G & 1.3 & 7 (0.4) & 2.54 (+1.16, -0.90)$^\sharp$ & 10 & 2.1 &  & Y\\
13 & 5$^\mathparagraph$ & 10$^h$ 58$^m$ 33.$^s$3 +36$\deg$ 12$^\prime$ 32$^{\prime\prime}$ & 8.9 & G & 1.4 & 7 (0.8) & 2.12 (+1.03, -0.80)$^\sharp$ & 10 & 2.9 &  & Y\\
14 & 1 & 09$^h$ 38$^m$ 14.$^s$9 +02$\deg$ 00$^\prime$ 23$^{\prime\prime}$ & 4.4 & G & 1.5 & 20 (0.5) & 7.3 ($\pm$3.6); 5.0 ($\pm$1.3) &  & 0.5 & 0.26 & Y\\
14 & 2$^\mathparagraph$ & 09$^h$ 37$^m$ 08.$^s$6 +01$\deg$ 25$^\prime$ 50$^{\prime\prime}$ & 6.4 & G & 1.5 & 7 (0.4) & 2.06 ($\pm$0.94) &  & 2.9 &  & Y\\
14 & 3$^\mathparagraph$ & 09$^h$ 37$^m$ 28.$^s$4 +01$\deg$ 35$^\prime$ 12$^{\prime\prime}$ & 6.1 & G & 2.1 & 8 (0.9) &  3.3 ($\pm$1.7) & 7.9 & 1.4 &  & \\
14 & 4 & 09$^h$ 39$^m$ 09.$^s$2 +01$\deg$ 44$^\prime$ 38$^{\prime\prime}$ & 5.1 & G & 1.4 & 14 (0.6) & 4.4 ($\pm$1.2) & 9.6 & 0.8 &  & Y\\
14 & 5$^\mathparagraph$ & 09$^h$ 37$^m$ 46.$^s$5 +02$\deg$ 08$^\prime$ 04$^{\prime\prime}$ & 5.9 & G & 1.6 & 7 (0.4) & 2.03 (+0.93, -0.72)$^\sharp$ & 12 & 2.9 &  & \\
15 & 1 & 18$^h$ 57$^m$ 53.$^s$8 +02$\deg$ 40$^\prime$ 08$^{\prime\prime}$ & 5.2 & G & 2.4 & 20 (1.3) & 13.0 ($\pm$7.3); 9.7 ($\pm$2.6) & & 1.3 & 0.47 & Y\\
15 & 2$^\mathparagraph$ & 18$^h$ 57$^m$ 41.$^s$1 +02$\deg$ 42$^\prime$ 07$^{\prime\prime}$ & 4.2$^{\dagger}$ & G & 2.1 & 13 (1.2) & 2.93 ($\pm$0.99) &  & 1.7 &  & Y\\
15 & 3$^\mathparagraph$ & 18$^h$ 58$^m$ 09.$^s$8 +02$\deg$ 21$^\prime$ 27$^{\prime\prime}$ & 4.8 & G & 0.5 & 12 (0.2) & 12.4 ($\pm$8.9) &  & 0.2 &  & Y\\
15 & 4$^\mathparagraph$ & 18$^h$ 59$^m$ 35.$^s$8 +02$\deg$ 42$^\prime$ 01$^{\prime\prime}$ & 7.3 & G & 1.6 & 7 (0.8) & 2.13 ($\pm$0.94) & 7.2 & 2.7 &  & \\
16 & 1$^\mathparagraph$ & 12$^h$ 47$^m$ 06.$^s$6 +15$\deg$ 12$^\prime$ 37$^{\prime\prime}$ & 4.2$^{\dagger}$ & G & 1.4 & 11 (0.7)  & 3.6 (+1.3, -1.1)$^\sharp$ & 12 & 1.1 &  & Y\\
16 & 2$^\mathparagraph$ & 12$^h$ 45$^m$ 46.$^s$9 +14$\deg$ 36$^\prime$ 27$^{\prime\prime}$ & 5.7$^{\dagger}$ & G & 0.7 & 9 (0.4) & 5.9 (+2.3, -1.9)$^\sharp$ & 11 & 0.6 &  & \\
16 & 3$^\mathparagraph$ & 12$^h$ 45$^m$ 37.$^s$6 +14$\deg$ 48$^\prime$ 57$^{\prime\prime}$ & 4.8$^{\dagger}$ & G & 0.8 & 10 (0.4) & 5.6 (+2.1, -1.7)$^\sharp$ & 15 & 0.6 &  & Y\\
16 & 4$^\mathparagraph$ & 12$^h$ 45$^m$ 35.$^s$6 +14$\deg$ 40$^\prime$ 11$^{\prime\prime}$ & 5.3$^{\dagger}$ & G & 0.8 & 10 (0.4) & 6.0 (+2.2, -1.8)$^\sharp$ & 13 & 0.6 &  & Y\\
16 & 5$^\mathparagraph$ & 12$^h$ 45$^m$ 37.$^s$9 +14$\deg$ 56$^\prime$ 36$^{\prime\prime}$ & 5.1$^{\dagger}$ & G & 1.5 & 10 (0.6) & 2.97 (+1.11, -0.90)$^\sharp$ & 11 & 1.3 &  & Y\\
18 & 1 & 19$^h$ 55$^m$ 03.$^s$6 +45$\deg$ 37$^\prime$ 04$^{\prime\prime}$ & 5.3 & G & 1.6 & 9 (0.6) & 3.4 ($\pm$1.2) & 6.4 & 1.6 &  & \\
18 & 2$^\mathparagraph$ & 19$^h$ 58$^m$ 25.$^s$8 +45$\deg$ 23$^\prime$ 29$^{\prime\prime}$ & 11.0 & G & 1.5 & 8 (0.5)  & 2.4 ($\pm$1.1) &  & 2.4 &  & Y\\
18 & 3$^\mathparagraph$ & 19$^h$ 58$^m$ 01.$^s$7 +45$\deg$ 18$^\prime$ 06$^{\prime\prime}$ & 5.6 & G & 1.6 & 7 (0.7) & 2.4 ($\pm$1.0) &  & 2.4 &  & Y\\
18 & 4$^\mathparagraph$ & 19$^h$ 54$^m$ 45.$^s$1 +45$\deg$ 43$^\prime$ 43$^{\prime\prime}$ & 5.1 & G & 1.4 & 7 (0.6) & 2.7 ($\pm$1.1) & 8.8 & 1.9 &  & \\
18 & 5$^\mathparagraph$ & 19$^h$ 59$^m$ 28.$^s$8 +45$\deg$ 17$^\prime$ 43$^{\prime\prime}$ & 7.8 & G & 1.5 & 8 (0.6) & 2.5 ($\pm$1.1) & 8.4 & 2.5 &  & \\
18 & 6$^\mathparagraph$ & 19$^h$ 58$^m$ 07.$^s$6 +45$\deg$ 35$^\prime$ 49$^{\prime\prime}$ & 4.9 & G & 1.6 & 10 (0.5) & 2.43 ($\pm$0.97) & 7.9 & 2.3 &  & \\
18 & 7$^\mathparagraph$ & 19$^h$ 58$^m$ 39.$^s$9 +46$\deg$ 01$^\prime$ 38$^{\prime\prime}$ & 7.0 & G & 1.8 & 7 (0.6) & 1.54 ($\pm$0.78) & 6.6 & 3.5 &  & \\
18 & 8 & 19$^h$ 58$^m$ 41.$^s$5 +45$\deg$ 34$^\prime$ 11$^{\prime\prime}$ & 9.7 & P & 3.0 & 5 (0.2) & 0.74 ($\pm$0.41) & 9.6 & 0.04 &  & \\
19 & 1 & 10$^h$ 21$^m$ 00.$^s$7 +16$\deg$ 25$^\prime$ 54$^{\prime\prime}$ & 4.7 & G & 1.3 & 24 (0.6) & 5.5 ($\pm$1.3); 5.1 ($\pm$1.1) &  & 0.3 & 0.33 & Y\\
19 & 2 & 10$^h$ 21$^m$ 28.$^s$6 +17$\deg$ 12$^\prime$ 44$^{\prime\prime}$ & 4.7 & G & 2.4 & 16 (1.0) & 1.97 ($\pm$0.52) & 12 & 1.3 &  & Y\\
19 & 3 & 10$^h$ 22$^m$ 49.$^s$1 +16$\deg$ 53$^\prime$ 45$^{\prime\prime}$ & 4.9 & G & 1.4 & 17 (0.5) & 7.4 ($\pm$1.9) & 26 & 0.6 &  & \\
19 & 4 & 10$^h$ 21$^m$ 51.$^s$5 +16$\deg$ 34$^\prime$ 32$^{\prime\prime}$ & 6.1 & G & 1.3 & 7 (0.5) & 4.6 ($\pm$2.0) & 44 & 2.1 &  & \\
19 & 5$^\mathparagraph$ & 10$^h$ 21$^m$ 18.$^s$2 +17$\deg$ 12$^\prime$ 29$^{\prime\prime}$ & 5.2 & G & 1.3 & 8 (0.4) & 3.2 ($\pm$1.2) & 24 & 1.5 &  & Y\\
19 & 6$^\mathparagraph$ & 10$^h$ 19$^m$ 35.$^s$0 +16$\deg$ 52$^\prime$ 54$^{\prime\prime}$ & 7.3 & G & 1.5 & 9 (0.6) & 2.71 (+1.09, -0.87)$^\sharp$ & 16 & 1.9 &  & \\
19 & 7$^\mathparagraph$ & 10$^h$ 21$^m$ 48.$^s$5 +17$\deg$ 03$^\prime$ 57$^{\prime\prime}$ & 8.7 & R & 2.4 & 7 (0.9) & 1.19 (+0.59, -0.46)$^\sharp$ & 8.3 & 1.1 &  & \\
20 & 1 & 17$^h$ 49$^m$ 35.$^s$9 +04$\deg$ 22$^\prime$ 29$^{\prime\prime}$ & 3.2$^*$ & G & 1.6 & 21 (1.2) & 3.3 ($\pm$1.4); 3.15 ($\pm$0.74) & 10 & 0.6 & 0.88 & Y\\
20 & 2 & 17$^h$ 50$^m$ 16.$^s$7 +04$\deg$ 22$^\prime$ 07$^{\prime\prime}$ & 5.3 & G & 1.6 & 9 (1.4) & 2.85 ($\pm$0.98) &  & 1.3 &  & Y\\
20 & 3$^\mathparagraph$ & 17$^h$ 47$^m$ 47.$^s$0 +04$\deg$ 53$^\prime$ 37$^{\prime\prime}$ & 5.7 & G & 0.9 & 7 (0.5)  & 3.0 ($\pm$1.3) & 14 & 1.5 &  & \\
\hline
\end{tabular}
\end{table}

\clearpage


\begin{table}
\caption{Details of the X-ray sources with spectra}
\label{tab:srcspec}
\begin{tabular}{ccccccc}
\hline
IceCube & Source \#   & $N_{\rm H,Gal}^a$  &  $N_{\rm H,intr}^b$   & Photon Index  &  ECF$^c$  & PSPC$^d$ \\
Field   &              & (\tim{20} \cms)    & (\tim{20} \cms)      &  & \tim{-11} erg \cms\ ct$^{-1}$ \\
\hline
1 & 1  &  27.77 &          $<$44.1                      & 1.95(+1.32, -0.70) & 3.4  & 1.11  \\ 
1 & 2  &  27.70 &          $<$260                       & $<$0.2 & 4600  & 0.01  \\ 
1 & 4  &  26.44 &          $<$87.9                      & 1.11(+1.40, -0.95) & 6.1  & 0.69  \\ 
3 & 2  &  0.93  &          $<$34.0                      & 1.06(+1.15, -0.80) & 5.7  & 1.18  \\ 
4 & 2  &  92.67 &          $<$37.6                      & 11.03(+2.81, -0.72) & 1.5  & 1.74  \\ 
5 & 7  &  3.35  &          $<$411                       & 2.4(+2.2, -1.3) & 2.4  & 1.43  \\ 
6 & 2  &  7.45  &          821(+1130, -790)             & 7.4(+26.5, -6.8) & 3.0  & 0.31  \\ 
6 & 3  &  6.64  &          $<$899.4                     & 1.1(+1.9, -1.0) & 5.7  & 0.85  \\ 
6 & 4  &  6.82  &          $<$48.1                      & 1.27(+1.01, -0.75) & 4.9  & 0.98  \\ 
6 & 7  &  8.23  &          $<$1004                      & 3.8(+3.2, -4.4) & 3.4  & 0.28  \\ 
6 & 10 &  6.54  &          $<$46                        & 2.38(+2.05, -0.90) & 2.6  & 1.45  \\ 
6 & 11 &  7.54  &          $<$69.4                      & 1.72(+2.26, -0.85) & 3.0  & 1.22  \\ 
7 & 1  &  21.64 &          $<$6.2                       & 3.39(+0.36, -0.19) & 1.9  & 1.65  \\ 
7 & 2  &  21.43 &          $<$56.0                      & 1.32(+1.53, -0.80) & 4.1  & 0.84  \\ 
7 & 4  &  21.20 &          $<$12.5                      & 1.96(+0.26, -0.20) & 3.6  & 1.15  \\ 
7 & 5  &  20.62 &          $<116$                       & 5.2(+12.9, -2.2) & 2.0  & 1.81  \\ 
7 & 6  &  21.13 &          $<$502                       & 3.0(+7.0, -2.7) & 2.3  & 1.01  \\ 
7 & 7  &  21.46 &          $<$85                        & 1.64(+1.04, -0.57) & 4.1  & 0.86  \\ 
7 & 8  &  21.31 &          240 (+546, -218)             & 2.05(+0.93, -0.85) & 6.7  & 0.24  \\ 
7 & 9  &  21.27 &          $<74$                        & 0.67(+1.14, -0.70) & 7.8  & 0.52  \\ 
7 & 10 &  20.99 &          $<$39.9                      & 2.30(+1.30, -0.55) & 2.3  & 1.34  \\ 
7 & 11 &  21.72 &          $<$44.7                      & 1.43(+1.25, -0.84) & 3.7  & 0.90  \\ 
7 & 12 &  21.48 &          $<$42.1                      & 2.6(+2.1, -1.0) & 2.5  & 1.44  \\ 
7 & 13 &  21.24 &          $<$355                       & 1.1(+2.0, -1.1) & 10.0  & 0.35  \\ 
7 & 14 &  21.43 &          $<$344                       & 2.2(+2.9, -1.0) & 5.1  & 0.48  \\ 
7 & 17 &  21.62 &          $<$305                       & 1.24(+2.82, -0.61) & 7.0  & 0.55  \\ 
7 & 18 &  21.30 &          $<$99.5                      & 1.37(+2.08, -0.92) & 4.3  & 0.87  \\ 
7 & 19 &  21.24 &          $<$95                        & 1.8(+1.7, -1.5) & 3.2  & 1.08  \\ 
7 & 20 &  21.26 &          $<$61.0                      & 1.50(+1.51, -0.79) & 4.0  & 0.94  \\ 
7 & 22 &  21.38 &          $<$89.1                      & 1.2(+1.9, -1.1) & 9.8  & 0.81  \\ 
7 & 23 &  20.97 &          541(1950, -225.0)            & 10.1(+82.8, -7.0) & 1.7  & 1.04  \\ 
7 & 30 &  21.56 &          39050(+10000, -8000)         & $>$50 & 2.6  & 1.50  \\ 
9 & 1  &  19.15 &          $<$200                       & 2.6(+3.4, -1.9) & 3.4  & 1.08  \\ 
9 & 3  &  19.01 &          $<497$                       & 5.4(+4.5, -4.8) & 2.0  & 1.10  \\ 
13 & 1  &  2.75  &          $<$28.8                      & 1.6(+1.3, -1.0)    & 3.9  & 1.49  \\ 
\hline
\end{tabular}
\medskip
\newline $^a$ The Galactic absorption value, taken from \protect{\cite{Willingale13}}.
\newline $^b$ The instrinsic absorption, fitted to the spectrum.
\newline $^c$ The conversion from XRT measured count-rate to 0.3--10 keV observed flux.
\newline $^d$ The conversion from XRT measured count-rate to that expected in the \rosat\ PSPC.
\end{table}        

\clearpage

\begin{table}
\contcaption{}
\begin{tabular}{ccccccc}
\hline
IceCube & Source \#   & $N_{\rm H,Gal}^a$  &  $N_{\rm H,intr}^b$   & Photon Index  &  ECF$^c$  & PSPC$^d$ \\
Field   &              & (\tim{20} \cms)    & (\tim{20} \cms)      &  & \tim{-11} erg \cms\ ct$^{-1}$ \\
\hline

14 & 1 &  3.44   &    $<$21.4                      & 2.37(+1.44, -0.54) & 3.1  & 2.10  \\ 
14 & 4 &  3.89   &    $<$34.0                      & 1.91(+2.13, -0.64) & 3.7  & 1.56  \\ 
15 & 1 &  144.00 &    $<$1312.4                    & 1.4(+3.3, -2.2) & 11.4  & 0.06  \\ 
18 & 1 &  26.20  &    $<$54.7                      & 1.63(+1.27, -0.88) & 4.3  & 0.96  \\ 
18 & 5 &  32.40  &    37200(6300, -13200)          & $^e$ & 4.4  & 1.50  \\ 
18 & 8 &  30.30  &    $<$476.9                     & 1.3(+3.4, -2.0) & 4.1  & 0.76  \\ 
19 & 1 &  3.11   &    $<$18.6                      & 2.40(+1.49, -0.39) & 2.4  & 2.27  \\ 
19 & 2 &  2.90   &    $<$93.5                      & 2.70(+4.96, -0.88) & 2.2  & 1.68  \\ 
19 & 3 &  2.90   &    $<$31.8                      & 1.33(+1.05, -0.56) & 4.9  & 1.18  \\ 
19 & 4 &  3.02   &    $<$274.9                     & 0.59(+1.42, -0.92) & 7.2  & 0.65  \\ 
20 & 1 &  16.55  &    $<$31.9                      & 2.84(+1.47, -0.78) & 2.1  & 1.55  \\ 
20 & 2 &  16.30  &    $<$79.9                      & 1.36(+1.15, -0.96) & 4.9  & 0.91  \\ 
\hline
\end{tabular}
\medskip
\newline $^e$ The photon index was unconstrained.
\end{table}

\clearpage
\end{landscape}

\begin{table*}
\begin{minipage}{166mm}
\caption{Details of the X-ray sources with catalogue matches}
\begin{tabular}{ccccc}
\hline
IceCube & Source \# & SIMBAD match & X-ray match & X-ray flux$^a$  \\
Field   &           & (arcsec) & (arcsec)  & (\tim{-13} \ergcms)  \\
       
\hline
3 & 1 & 2MASS J14590700+5931128 ( 2.9$^{\prime\prime}$) & 1RXS J145906.5+593105 (8.2$^{\prime\prime}$) & $6.9\pm2.4$\\
5 & 6 & [VV2010c] J083046.9+185707 ( 2.0$^{\prime\prime}$) &  & \\
6 & 6 &  & 1WGA J0125.4+1003 (55.6$^{\prime\prime}$) & $4.23\pm0.67$\\
7 & 1 & TYC   76-1038-1 ( 2.9$^{\prime\prime}$) & 1RXS J040840.9+033448 (9.7$^{\prime\prime}$) & $11.9\pm2.3$\\
7 & 3 & HD  26292 ( 2.5$^{\prime\prime}$) &  & \\
9 & 2 & V* V1296 Tau ( 3.7$^{\prime\prime}$) & 1RXS J040307.5+044427 (22.4$^{\prime\prime}$) & $8.9\pm3.2$\\
9 & 4 & TYC   79-810-1 ( 2.7$^{\prime\prime}$) & 1RXS J040428.2+051854 (10.9$^{\prime\prime}$) & $12.5\pm3.6$\\
10 & 1 & [VV2000] J223949.1+060613 ( 6.8$^{\prime\prime}$) & 1RXS J223949.7+060610 (13.0$^{\prime\prime}$) & $6.0\pm2.5$\\
11 & 2 & TYC  902-318-1 ( 4.3$^{\prime\prime}$) & 1RXS J141546.6+080755 (19.6$^{\prime\prime}$) & $17.9\pm4.6$\\
13 & 1 & MCG+06-24-038 ( 2.0$^{\prime\prime}$) &  & \\
13 & 3 & USNO-A2.0 1200-06616808 ( 3.7$^{\prime\prime}$) &  & \\
13 & 4 & SDSS J105631.92+354152.8 ( 0.8$^{\prime\prime}$) &  & \\
13 & 5 & SDSS J105833.38+361228.3 ( 4.2$^{\prime\prime}$) &  & \\
14 & 1 & 2MASX J09381483+0200236 ( 1.2$^{\prime\prime}$) & 1RXS J093815.6+020026 (10.8$^{\prime\prime}$) & $5.0\pm1.6$\\
14 & 2 & 2MASS J09370850+0125440 ( 6.2$^{\prime\prime}$) & 1RXS J093708.2+012620 (31.1$^{\prime\prime}$) & $5.5\pm2.2$\\
14 & 4 & SDSS J093909.42+014433.5 ( 5.7$^{\prime\prime}$) &  & \\
15 & 1 &  & 1SXPS J185753.6+024012 (5.0$^{\prime\prime}$) & $9.4\pm1.6$ \\
15 & 2 & BD+02  3740 ( 4.1$^{\prime\prime}$) & 1SXPS J185741.3+024208 (2.6$^{\prime\prime}$) & $2.40\pm0.45$\\
15 & 3 &  & 1SXPS J185809.7+022131 (3.9$^{\prime\prime}$) & 5.4 (+1.4, -1.2)\\
16 & 1 & LBQS 1244+1529 ( 2.9$^{\prime\prime}$) &  & \\
16 & 3 & 2MASS J12453751+1448572 ( 1.5$^{\prime\prime}$) &  & \\
16 & 4 & LBQS 1243+1456 ( 1.3$^{\prime\prime}$) &  & \\
16 & 5 & [VV2006] J124537.7+145635 ( 4.2$^{\prime\prime}$) &  & \\
18 & 2 & 1RXS J195826.1+452321 ( 8.7$^{\prime\prime}$) & 3XMM J195825.7+452325 (3.3$^{\prime\prime}$) & $1.467\pm0.048$\\
18 & 3 & 2MASS J19580115+4518060 ( 6.1$^{\prime\prime}$) & 3XMM J195801.1+451805 (6.3$^{\prime\prime}$) & $0.637\pm0.017$\\
19 & 1 & SDSS J102100.35+162554.0 ( 5.0$^{\prime\prime}$) & XMMSL1 J102100.3+162550 (5.5$^{\prime\prime}$) & $37.8\pm9.6$\\
19 & 2 & LP  430-7 ( 3.3$^{\prime\prime}$) &  & \\
19 & 5 & SDSS J102118.34+171227.7 ( 3.7$^{\prime\prime}$) &  & \\
20 & 1 & HD 162178 ( 8.9$^{\prime\prime}$) &  & \\
20 & 2 &  & XMMSL1 J175016.8+042202 (5.9$^{\prime\prime}$) & 35$\pm$11 \\
\hline
\end{tabular}
\medskip
\newline $^a$ The catalogued 0.3--10 keV flux. For \rosat\ objects, this is the catalogued count-rate, converted
to 0.3--10 keV flux using the conversion from Table~\ref{tab:sources}; for other sources it is the catalogued flux, which is 
already in the 0.3--10 keV band (1SXPS sources) or the similar 0.2--12 keV band (\xmm\ sources).
\label{tab:catmatch}
\end{minipage}
\end{table*}

\section{Identifying potential counterparts to the neutrino trigger}
\label{sec:counterpart}

Each 7-tile observation with \swift-XRT covers \til0.8 square degrees, with typical exposures of 1--2 ks per tile.
In such exposure times, our detection limit is 6--10\tim{-13} \ergcms [corresponding to 90\%\ completeness; see \cite{Evans14}, fig. 14].
This is significantly more sensitive than the RASS, which covers 92\%\ of the
sky down to 0.1 ct s$^{-1}$ in the PSPC\footnote{For individual fields the RASS limit may be deeper.}
\citep{Voges99}, which corresponds to 2.8\tim{-12} \ergcms\ in the 0.3--10 keV 
band covered by XRT (assuming the canonical
AGN spectrum described above). For absorbed sources (i.e.\ with little flux in the \rosat\ bandpass) the increase in sensitivity
of \swift-XRT over \rosat\ is even greater.
The XSS provides hard-band coverage, being sensitive to \til3\tim{-12} \ergcms\ in the 2--10 keV band \citep{Warwick12}, however it only covers
\til2/3 of the sky. Because of the sensitivity of the XRT compared to these catalogues, and the low spatial coverage of 
deeper catalogues like 1CSC \citep{ievans10}\footnote{No relation.}, 3XMMi-DR4
(Watson et al., in preparation) and 1SXPS \citep{Evans14}, we expect to discover uncatalogued X-ray sources
that are serendipitously present in the IceCube error region. We therefore need to be able to identify the true
X-ray counterpart to the neutrino trigger from among the unrelated objects detected in the field of view.
The first step is to remove any sources which are already known X-ray emitters (and are not in outburst
at the time of the \swift\ observations). We searched the X-ray Master catalogue\footnote{http://heasarc.gsfc.nasa.gov/W3Browse/master-catalog/xray.html\\
http://ledas-www.star.le.ac.uk/arnie5/arnie5.php?action=basic\\
\&catname=xcoll} for any catalogued X-ray object with a position agreeing with 
our XRT position at the 3-$\sigma$ level (including any systematic errors on the catalogued positions). For all observations after 2012 October\footnote{i.e.\ the
dates covered by the 1SXPS catalogue; the observataions of IceCube fields prior to this date are in the 1SXPS catalogue, therefore
it cannot be used as a reference in those cases.} we also searched for matches in the 1SXPS catalogue. The sources with catalogue matches are indicated in
Table~\ref{tab:sources}, and details of the matches are given in Table~\ref{tab:catmatch}.
In all cases where a match was found, the XRT flux was consistent with (at the 2-$\sigma$ level) or occasionally slightly lower than the catalogued flux, 
therefore these sources were all rejected as possible counterparts.

For the remaining, uncatalogued sources we performed two tests to identify the counterpart: variability tests, and serendipity likelihoods.

\subsection{Variability}
\label{sec:counterpart}

\begin{figure}
\begin{center}
\psfig{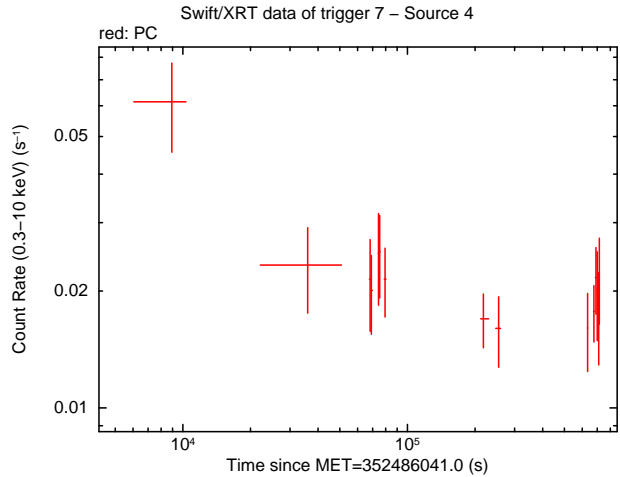}
\end{center}
\caption{Source 4 of the 7th IceCube trigger. This source momentarily peaked above the
RASS upper limit for its location, and was fading; however further observations showed the
brightness to have levelled off. This is probably a background AGN.}
\label{fig:notfading}
\end{figure}

The simplest test of whether an uncatalogued X-ray source is likely to be related to the neutrino trigger is 
to ask whether the source brightness is such that is should have been seen previously, i.e.\ the measured brightness
is significantly above the sensitivity limit of any instruments which have previously observed the source location
without discovering the source. If this is the case then the source is in a bright state compared
to previous observations, and is thus a good candidate to be the counterpart.

For each uncatalogued source, we therefore derived a 3-$\sigma$ upper limit on the \rosat\ PSPC count rate 
using images, background maps and exposure maps obtained from ftp://ftp.xray.mpe.mpg.de/ftp/rosat/archive, and the Bayesian method
of \cite{Kraft91}. Where available, we also obtained 3-$\sigma$ upper limits on the 0.2--12 keV flux from the XSS via the 
service at http://xmm.esac.esa.int/UpperLimitsServer\footnote{This service also provides upper limits
derived from pointed \xmm\ observations, however with the exception of the two sources, which have been previously
detected by \xmm\, none of the objects found by XRT have been in the field of pointed \xmm\ observations.}
In no cases was the mean XRT brightness above these upper limits at the 1-$\sigma$ level, although in one source (discussed below) the flux did briefly peak above the RASS upper limit.
The upper limits for each detected source are given in Table~\ref{tab:sources}.

A second test is to see if the source is variable during the \swift\ observations, particularly whether it is showing
significant signs of fading. Such behaviour is expected from explosive transients such as GRBs or SNe; Tidal Disruption Events (TDEs) also fade, although 
they may undergo a period of \til constant flux for some time before fading \citep{Lodato11}.
Following \cite{Evans14} we calculated the Pearson's \chisq\ for each source where we were able to produce a light curve (i.e.\ with more than
one bin), using as a model a constant source with count-rate set to the mean detected rate for that source. The results of this test
are shown in Table~\ref{tab:sources}. The drawback to this method is that it requires binned data,
therefore one could argue that the K-S test would be a better indicator of variability.
However, many objects show flickering behaviour in X-rays (e.g.\ CVs) which is not indicative of ourbursting. We therefore 
require not only a measure of variability, but also a light curve that we can inspect, after being prompted
by that measure. This implicitly demands binned data, hence our choice of the \chisq\ test.

Only one uncatalogued object (out of 14 for which we can probe 
variability) has a probability of $<10\%$ of being constant, source 4 of 
field 7; this is the same source which briefly peaked above the RASS 
upper limit. In the initial observations, the light curve of this source 
comprised 2 bins, with evidence for fading at the 2.3-$\sigma$ level. We 
therefore observed the source again repeatedly for a week, but no signs 
of continued fading were seen (Fig.~\ref{fig:notfading}). This source, 
which is spatially coincident with the USNO-B1 source 0935-0049746, is 
therefore consistent in behaviour with a background AGN, from which 
variations of factors of \til2 are not uncommon; there is no evidence for
a powerful flare that could have produced a high neutrino flux (this would
have yielded a much stronger EM flare than that observed). Detailed examination of 
the IceCube data showed that the event was fully consistent with 
detection of two neutrinos (i.e.\ not cosmic-ray-induced muons), but the 
significance was at the level at which \til4 false positives are 
expected per year. As a final check of this source we re-observed it 
with \swift-XRT on 2014 September 17. In this observation the source was 
still easily detected, with a count-rate of \til0.03 ct s$^{-1}$, 
consistent with the previous observations and further arguing against 
its being related to the IceCube trigger.

One other source (source 1 in field 7) also showed strong
signs of variability ($P_{\rm constant}=6\tim{-4}$), however this was a catalogued X-ray source, 1RXS J040840.9+033448, which SIMBAD\footnote{http://simbad.u-strasbg.fr/simbad/}
lists as a rotationally variable star, and it was a factor of two fainter in our observations than in the \rosat\ catalogue, therefore we
do not consider this to be a probable counterpart to the neutrino trigger.

\subsection{Probability of serendipity}
\label{sec:pseren}

If we can identify a given source as being unlikely to be serendipitously present in the field
of view then it is, conversely, a strong candidate to be the counterpart to the neutrino trigger and
worthy of further observation. A simple metric to use
to quantify this likelihood is the source brightness. 
To determine the probability of serendipity for a given source, we first determined the minimum exposure necessary to detect the source:
for our detection system we require at least 6 counts, therefore the minimum exposure
is $6/R$ (where $R$ is the XRT count-rate of the source). We then measured the  sky area, $A$, in our tiled observation
which was observed with at least this exposure (accounting for vignetting, overlaps etc.). We
used the log $N$-log $S$ brightness distribution of extra-galactic sources, calculated by 
\cite{Mateos08} (based on the data in the 2XMM catalogue, \citealt{Watson09}) to determine the expected
sky density of sources, $D$ at least as bright as the detected object in question. The number
of expected serendipitous sources in our field of view, at least as bright as the detected object, is $A\cdot D$.
This was then multiplied by the completeness of our detection system for the source brightness and exposure time to
account for the fact that not all such sources will have been detected. This yields the number
of serendipirous sources at least as bright as the detected source that we would expect to detect
in our observations. This value is given, for each source, in Table~\ref{tab:sources}. 

Only one uncatalogued source (field 18, source 8) was found with a low probability of serendipity. This source is faint, but lies 
in a region covered by three of the tiled pointings and therefore with a deeper exposure than most of the image. The sky area covered
with such an exposure is very low, which is driving the low probability of serendipity. However, the source is flagged as \emph{Poor} by
the detection system, meaning that it has a 35\%\ probability of being a spurious detection, and the image does not show a strong clustering of
events as expected from a point-like source, therefore we think this source is unlikely to be real.

\begin{figure}
\begin{center}
\psfig{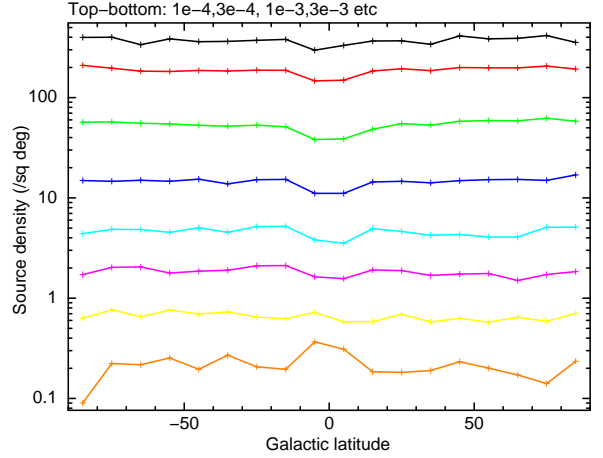}
\end{center}
\caption{The density of X-ray objects above a given flux as a function of Galactic latitude. 
For bright sources there is a slight increase in the density towards the Galactic centre, and for fainter sources
there is a slight deficit, however the effect is small. For a typical AGN spectrum 1 XRT count corresponds
to 4\tim{-11} erg cm$^{-2}$. From top to bottom, the lines correspond to $10^{-4}, 3\tim{-4},10^{-3}, 3\tim{-3},
10^{-2}, 3\tim{-2},10^{-1}, 3\tim{-1}$ counts per second.}
\label{fig:latdep}
\end{figure}

The log $N$-log $S$ curve of \cite{Mateos08} was derived only for 
Galactic latitudes $|b|>20\deg$, whereas some of the \ic\ triggers were 
at lower Galactic latitudes. We therefore used the 1SXPS catalogue 
\citep{Evans14} to investigate the Galactic dependence of the X-ray 
source density. For each field in the 1SXPS catalogue we calculated the 
density of sources brighter than $x$ ct s$^{-1}$, for values of 
$x=1\tim{-4}, 3\tim{-4},1\tim{-3}, 3\tim{-3},1\tim{-2}, 
3\tim{-2},1\tim{-1}, 3\tim{-1}$. We did this using the {\sc healpix} 
libraries \citep{Gorski05} with {\sc nside}=8192, which results in 
pixels of area 0.18 square arcmin. Since \swift\ observes many fields 
multiple times, for each field in 1SXPS we determined for each {\sc 
healpix} pixel which observation or stacked image contained the most 
exposure. Only that dataset and sources detected in it, were considered. 
For field $M$, the density of sources brighter than $x$ is:

\begin{equation}
\Omega_{M, >x} = \sum_N{ \frac{1}{A_N C_N}}
\end{equation}

\noindent where the sum is over all $N$ sources in the field, $A_N$ is the area (in the {\sc healpix} map)
covered by the observation in which source $N$ was detected, and $C_N$ is the completeness
of the 1SXPS catalogue for sources at least as bright as source $N$ (i.e.\ for each source we detect, there are $1/C_N$ actual sources). 
The overall density as a function of Galactic latitude can
then be found simply as:

\begin{equation}
\Omega_{b, >s} =  \frac{\sum_M{\Omega_{M,>s} A_M}}{\sum_M{A_M}}
\end{equation}

\noindent where the sum is over all fields $M$ with Galactic latitude $b$. We compared our results for $|b|>20\deg$ 
with \cite{Mateos08}, and found that at XRT count rates below \til 2\tim{-3} ct s$^{-1}$ (equivalent to 0.3--10 keV flux \til6.3\tim{-14} \ergcms)
they were in good agreement. At higher count-rates we predicted slightly more sources than \cite{Mateos08}, probably because we did not 
remove from our source list the objects which were the target of the \swift\ observations, whereas \cite{Mateos08} did, meaning our results are slightly biased.
Nonetheless, we can use our results to give an indication of the Galactic latitude dependence of source density, so we
split the data into bins of 5\deg\ in Galactic latitude; the source density as a function of latitude and count-rate is shown
in Fig.~\ref{fig:latdep}. There is a small effect seen towards the Galactic plane: at most source brightnesses there is
a slight reduction in source density at $|b|<10\deg$, presumably related to the increased absorption in the Galactic plane, however
this reduction is less than \til10\%. At the brightest fluxes there is an increase of \til40\%\ in the density of sources
at $|b|<10\deg$, however this is likely to be effected by pointed observations of transient objects, i.e.\ this
does not accurately reflect the density of bright serendipitous sources in the Galactic plane. Based on this analysis, we believe 
that it is safe to use the extra-Galactic log $N$-log $S$ distribution of \cite{Mateos08} for all of the IceCube triggers, regardless
of Galactic latitude.

\section{Discussion and implications for multi-messenger astronomy}
\label{sec:disc}

We have reported on \swift-XRT X-ray follow up of 20 neutrino triggers from the IceCube observatory.
Although we have found 109 X-ray sources, none of them have been identified as a likely counterpart to an IceCube trigger.
Conversely, only the 30 objects listed in Table~\ref{tab:catmatch} were known prior to the \swift\ observations, and only 16 of them had been 
detected in X-rays previously, therefore it is very difficult for us to rule out with any degree of confidence that
we have detected an electromagnetic (EM) counterpart to the neutrino trigger, even though we cannot identify it.
To investigate further we therefore need to consider what we expect the counterparts to look like in our follow-up observations.
Also, IceCube continues to send neutrino triggers to \swift\ and the field of multi-messenger astronomy 
at large is growing rapidly: for example, advanced LIGO (aLIGO) is expected to be commissioned in early 2015 \citep{Harry10}.
Therefore, it is necessary to consider whether lessons can be learned from the IceCube followup presented in this paper, for future
EM follow up on non-EM triggers. Note that we will focus on 
how to improve our ability to identify the counterpart from among the sources detected in an observation: we do not
consider the related issue of how optimally to observe a large (and potentially disjoint) error 
region; for a discussion of that subject, see \cite{Kanner12}.

Although there are a range of phenomena that can produce neutrino triggers, in this discussion we
focus on GRBs (which are also a potential source of gravitational waves), since these have been well studied in the
X-ray domain by \swift\ and, as the brightest known transients in the universe, they represent an upper bound on the 
detectability of the EM counterparts to neutrino triggers.


\subsection{How bright will the X-ray counterpart be?}
\label{sec:discBright}

\begin{figure}
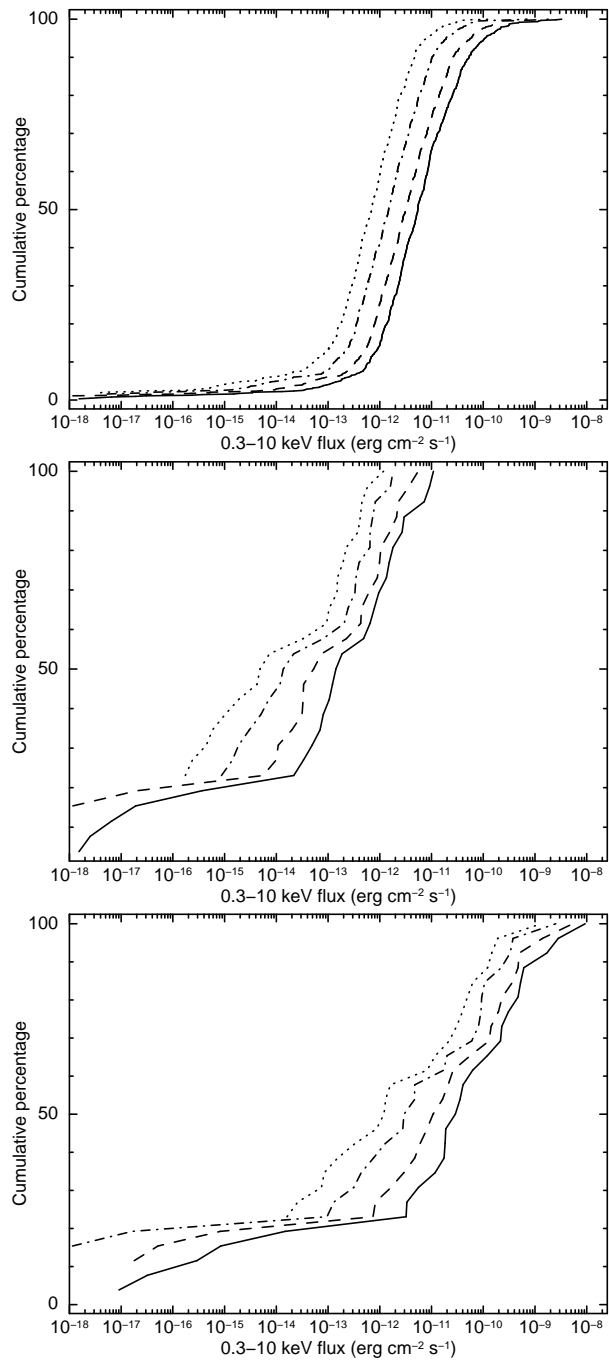

\begin{center}
\psfig{file=fig6a.eps,height=8.1cm,angle=-90}
\psfig{file=fig6b.eps,height=8.1cm,angle=-90}
\psfig{file=fig6c.eps,height=8.1cm,angle=-90}
\end{center}
\caption{The cumulative distribution of the brightness of GRB afterglows detected by the \swift-XRT. The solid line
showed the flux 1 hour after the trigger, the dashed line is 1.8 hours after the trigger, and the dot-dashed and dotted lines are for 4 and 8 hours after 
the trigger, respectively. The brightness is taken by evaluating the best-fitting model at these times, which may 
involve extrapolating past the time a GRB ceases to be detected by \swift.
The top panel shows all GRBs, the centre panel shows only the short GRBs. The bottom panel is as the centre panel,
but with the GRBs shifted to be at a distance of 200 Mpc [assuming the redshifts given in \protect\cite{Rowlinson13}].}
\label{fig:grbs}
\end{figure}

\begin{figure}
\begin{center}
\psfig{file=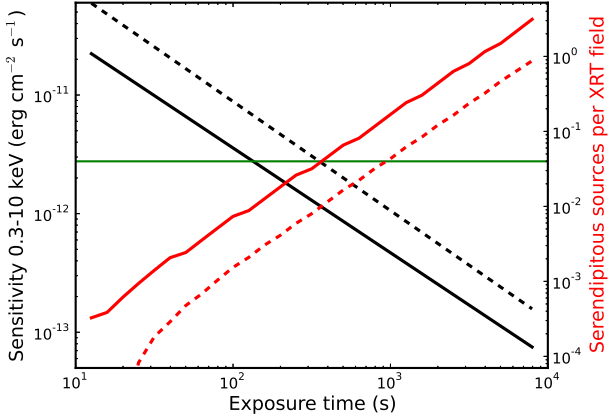,width=8.1cm}
\end{center}
\caption{The sensitivity of \swift-XRT (black lines), and the expected 
number of serendipitous source expected per XRT field above this limit 
(red lines), as a function of exposure time. The solid lines correspond to the 50\%\ completeness 
level, and the dashed lines the 90\%\ completeness level. Note that the two $y$-axes do not correspond with each
other, but are only related via the $x$-axis and plotted data.
\newline The green horizontal line shows the sensitivity limit of the RASS \protect\cite{Voges99}, corresponding to 0.1 
PSPC ct s$^{-1}$ (converted to 0.3--10 keV flux assuming the canonical 
AGN spectrum described in the text), at which level the RASS covers 
92\%\ of the sky. The \xmm\ slew survey 2--10 keV band limit is at a 
similar level (3\tim{-12} \ergcms; \protect\citealt{Warwick12}).}
\label{fig:sensseren}
\end{figure}

In our analysis (Section~\ref{sec:counterpart}) we looked for, and failed to find, any uncatalogued sources in our
dataset which were so bright that they would have been catalogued if they were persistent. The lack of such sources
may indicate that we did not find the counterpart to the IceCube trigger, or simply that we observed too late after the trigger,
when the source had faded below the limit of the existing catalogues.
How bright the EM counterpart to a non-EM trigger is at a given time depends on the 
nature of the emitting object. Here we consider GRBs \citep{Klebesadel73}, since they are 
expected to be sources of neutrinos and gravitational waves. GRBs are the brightest known EM transients in the universe,
and, although they fade fairly rapidly, they give us a reasonable upper limit on the brightness
we may expect from the X-ray counterpart to a non-EM trigger.

The top panel of Fig.~\ref{fig:grbs} shows the distribution of the X-ray flux of GRB afterglows observed with XRT,
at a range of times since the trigger. These were derived using the live XRT GRB catalogue\footnote{http://www.swift.ac.uk/xrt\_live\_cat}
\citep{Evans09} from which we used the light-curve fits for all GRBs with at least 5 light-curve bins (i.e.\
where the fit was reasonably well constrained), and an ECF of 4.1\tim{11} erg \cms\ ct$^{-1}$, which is the mean value
in that catalogue. About 60\%\ of GRBs are expected to be above the the typical RASS/XSS sensitivity limit of \til3\tim{-12} \ergcms\ (0.3--10 keV)
at one hour after the trigger, this falls to about 15\%\ by eight hours. Comparing this with the delay between the IceCube trigger
and the start of the \swift\ observations (Table~\ref{tab:icobs}) we would expect that in \til8 of cases we should have found an 
uncatalogued source above the RASS/XSS limit, if the neutrino triggers were related to GRBs. The lack of any such object
rules out the idea that all 20 triggers arose from GRBs with $>$99\%\ (i.e.\ 3-$\sigma$) confidence. However, the companion paper to this one
(Aartsen et al., in preparation) shows that many (or all) of the neutrino triggers could have been spurious;
if even half of the triggers were spurious, this significance drops to below 3-$\sigma$.

The lack of bright sources does not mean that we did not detect a GRB afterglow: in more than half of the triggers,
by the time \swift\ observed, the afterglow would have faded below the RASS/XSS limit. However, the ability
to identify an afterglow at these lower fluxes is hampered by the density of expected (uncatalogued) sources,
as illustrated in Fig.~\ref{fig:sensseren}. This shows the level (black) at which XRT is 50\%\ and 90\%\ complete
\citep{Evans14}, and the expected number of serendipitous sources (red) per XRT field of view above these levels
(Section~\ref{sec:pseren}) as a function of exposure time. The green line corresponds to the typical RASS/XSS limit.
The XRT 90\%\ completeness level reaches the RASS and XSS limits in an exposure of \til350 s; and we 
expect \til0.01 serendipitous sources per XRT field with fluxes above this limit. That is, in a 7-tile observation
such as those reported in this paper, any detected source below the flux limit set by the existing large-area catalogues,
will have a probability of being serendipitous of $\geq$0.07, i.e.\ we cannot expect to identify the counterpart with
even 2-$\sigma$ confidence.

It is impossible therefore, for us to identify the counterpart to the neutrino triggers reported in this paper 
based on the source flux at detection, and in any future follow-up of astrophysical neutrinos, we would expect at best 50\%
of GRB afterglows to be identified in this way. 

While neutrinos are expected from all GRBs, a prime candidate for the sources of gravitational
waves are nearby short GRBs, which arise from the merging of two neutron stars.
The middle panel of Fig.~\ref{fig:grbs} shows the flux distribution of the short GRBs detected
by the \swift-XRT: they are much fainter than long GRBs and we are unlikely to observe any before they 
fall below the limits of existing catalogues. However, the horizon distance of aLIGO is around 200 Mpc \citep{Abadie10},
whereas the  the average short GRB redshift in the \swift\ sample is 0.72 \citep{Rowlinson13}, 
corresponding to a luminosity distance of \til4000 Mpc. Thus on-axis short GRBs detected by 
aLIGO should be a factor of \til400 brighter than those detected by \swift, although the time-axis
of the light curve is compressed by the reduced time dilation, which shortens any plateau phase. In the bottom panel of Fig.~\ref{fig:grbs},
we have shifted the XRT afterglows from the redshifts given in \cite{Rowlinson13} to 200 Mpc ($z=0.045$). In this case \til 80\%\ of short GRBs
would be above the RASS limit one hour after the trigger, and 50\%\ would still be that bright at eight 
hours. These results are less optimistic than those reported by \cite{Kanner12}, however they used only short GRBs with 
known redshift (giving a smaller sample), whereas we have included short GRBs with no known redshift, assigning to them
the mean short GRB redshift of 0.72. It should also be noted, that in ten years of operation, \swift\ has not yet detected a short GRB less
than 500 Mpc away (GRB 061201, $z=0.111$ \citealt{Berger06a}), and indeed no short GRB thousands of times brighter than the typical \swift\
short GRBs has been reported in over twenty years of observations by various facilities. This tells us that nearby short GRBs,
which may trigger aLIGO, are extremely rare.

\subsubsection{Increasing the sensitivity}
\label{sec:preImage}

Our ability to identify a counterpart by its brightness would be enhanced if we had a more sensitive reference catalogue.
For example, Fig.~\ref{fig:sensseren} shows
that if \swift-XRT had conducted a 2 ks observation of a field prior to an IceCube trigger, then the list of known sources 
at that location would be 90\%\ complete down to a flux 5 times below the RASS limit; for hard or absorbed sources the increase in sensitivity is 
significantly more pronounced. At such levels, 95\%\ (50\%) of the \swift-detected GRBs would be bright enough to
be confirmed as new (non-serendipitous) sources in an observation at 1 (8) hours after the trigger.

To pre-image the entire sky with \swift-XRT, at 2-ks per field, is clearly not practical (it would require 
around 18 years of observing time!), although some subset of the sky, for example, corresponding to the galaxies deemed most
likely to yield a short GRB that aLIGO would detect, could potentially be observed. The forthcoming \emph{eRosita}
mission, expected to launch in 2016, will produce an all-sky survey in the 0.2--10 keV band which will be
a factor of 30 more sensitive than the RASS \citep{Cappelluti11}. This will provide a valuable resource 
for identifying new sources in \swift\-XRT observations of non-EM triggers. In the meantime, catalogues such as 
the 1CSC \citep{ievans10}, 3XMMi-DR4 (Watson et al., in preparation) and 1SXPS \citep{Evans14} could be used when available,
but their sky coverage is very limited.

\subsection{Identifying counterparts by fading light curves}
\label{sec:fade}

\begin{figure}
\begin{center}
\psfig{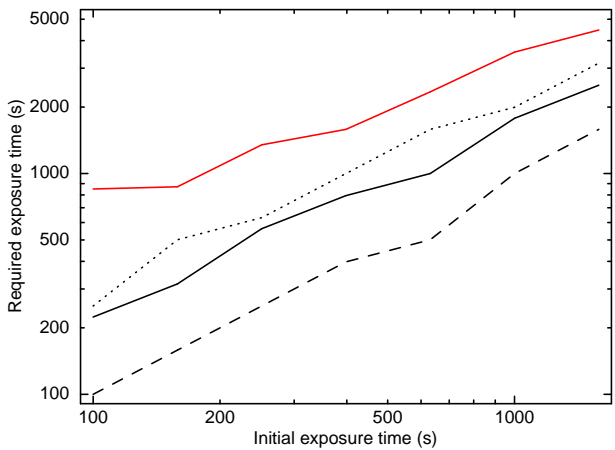}
\end{center}
\caption{The minimum second-observation exposure time needed to detect the fading of a source as a function of the exposure time
in the initial image. The source brightness in the first observation is that at which our detection system is 50\%\ complete in the initial exposure time.
If the source is detectable in the second observation, the black (red) solid line shows the minimum time needed to detect the source in the second image, 
and measure the count-rate as 2 (3) $\sigma$ fainter than in the initial image. 
If the source is not detectable in the second image, the black dotted line shows the exposure time needed to obtain
a 3-$\sigma$ upper limit inconsistent with the initial brightness if the source is only just below the detection threshold; the black
dashed line is that needed if the source has faded away completely. See Section~\ref{sec:fade} for full details.}
\label{fig:fadetest}
\end{figure}

Transient events by definition fade over time. However, in our follow-up observations, only 19 (out of 109) sources
were bright enough (or observed for long enough) to yield two or more light curve bins, and 12 of these occured in the field of trigger \#7,
which was observed for an unusually long time to allow us to rule out the possible counterpart in that field (Section~\ref{sec:counterpart}).
Also, not all transient sources fade on the timescale
of a single observation. GRBs, for example, tend to fade quickly, but many show a `plateau' phase where there is little or no
decay (\citealt{Zhang06,Nousek06}); tidal disruption events similarly begin with a period of roughly constant flux before
beginning their decay \citep{Lodato11}. To discover whether a source is fading it is therefore necessary to perform repeat observations,
and if the type of transient is not known, then several such observations are needed, 
with increasing delays since the trigger to account for different progenitor types. While this strategy was not
employed for the neutrino triggers reported in this paper, such an approach should be considered in the future, given the
lack of identified counterparts in this work.

\begin{figure*}
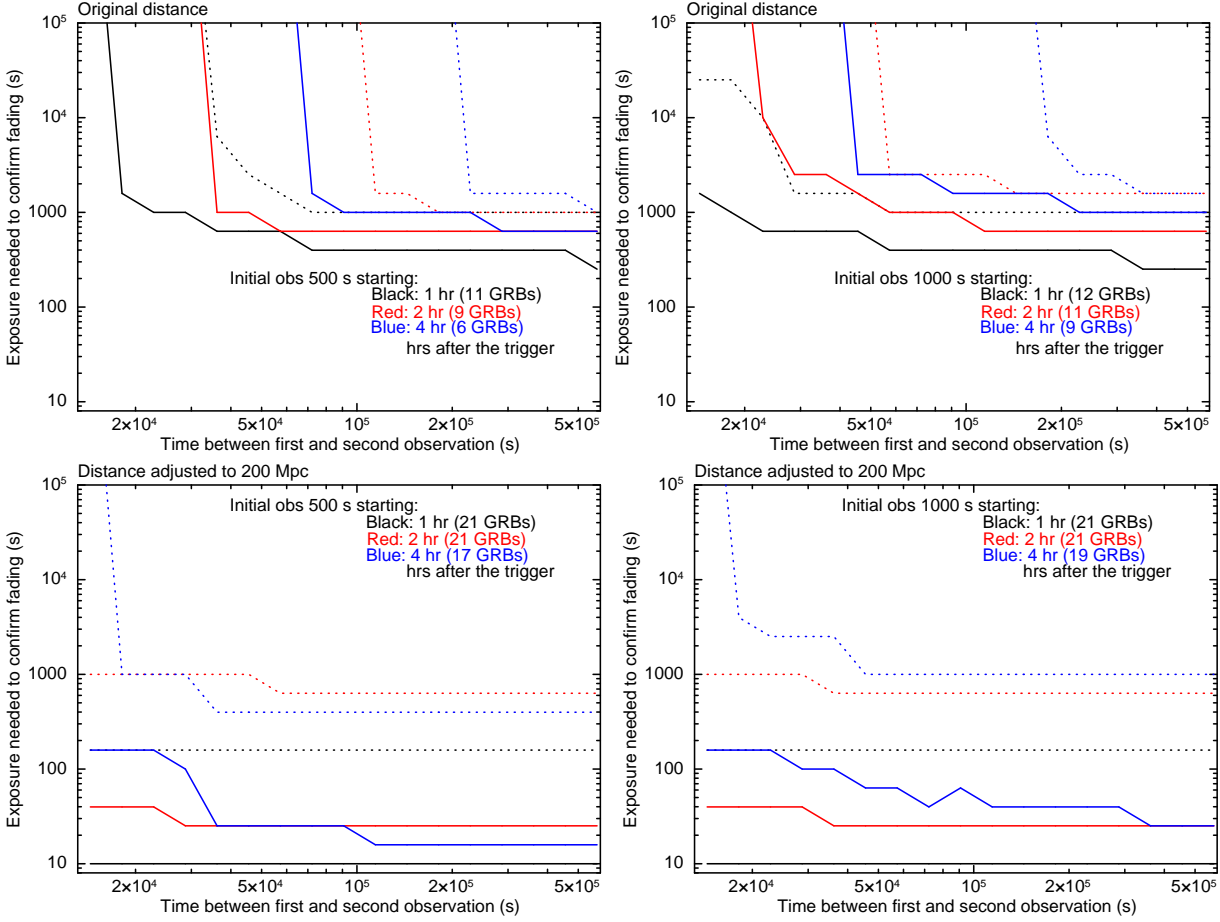

\begin{center}
\hbox{\psfig{file=fig8a.eps,height=8.1cm,angle=-90}\psfig{file=fig8b.eps,height=8.1cm,angle=-90}}
\hbox{\psfig{file=fig8c.eps,height=8.1cm,angle=-90} \psfig{file=fig8d.eps,height=8.1cm,angle=-90}}
\end{center}
\caption{The exposure time necessary in a follow-up observation of a short GRB needed to confirm fading at the 3-$\sigma$ level, as
a function of the delay between the initial and follow-up observations. The solid (dotted) lines are the exposure needed
for 50\%\ (90\%) of the short GRBs detected in the initial observation to be confirmed fading at the 3-$\sigma$ level.
The black, red and blue lines are for an initial observation beginning 1, 2 and 4 hours after the GRB trigger respectively. 
The numbers in the legend indicate the number of GRBs corresponding to 100\% for each dataset, i.e. the number of GRBs
that would have been detected in the initial observation. This is from a total of 43 GRBs which are input to the simulations.
\newline \emph{Left:} For an initial observation of 500 s. \emph{Right:} For an initial observation of 1000 s.
In the bottom panels we have adjusted the light curves of the short GRBs as if the GRB was at 200 Mpc, using the redshifts and GRB list given in \protect\cite{Rowlinson13}.
In the top panels, the GRB light curves as detected by \swift\ were used.
\newline The point where the lines stop decyaing reflects corresponds to the point at which the 50th or 90th percentile becomes too faint to detect,
i.e.\ to confirm fading we require an upper limit below the level of the initial detection. This level, and hence the time needed
to generate the upper limit, does not depend on the time between observations.}
\label{fig:grbfade}
\end{figure*}

In order to confirm fading, these extra observations need to be of sufficient exposure for us to
either measure the count-rate accurately enough to confirm a decay with some predetermined level of significance, or find an upper
limit below the level of the previous detection. To accurately determine how much exposure is needed requires knowledge of
the source light curve. As a generic approach, we can determine the count-rate $R_2$ in the second observation
which allows us to confirm fading in the shortest possible exposure time\footnote{This is a function of the
background level and the size of region over which source counts are accumulated. We set the background to $10^{-6}$ ct s$^{-1}$
pixel$^{-1}$, the mean value from the 1SXPS catalogue \citep{Evans14}, and set the region size to be that used by the XRT auto-analysis
software for the brightness of the source in the initial observation, see table~1 of \cite{Evans07}.}. Since the optimal value of $R_2$ is not
a priori obvious, we stepped it over the range  $0.001R_1$ -- $0.99R_1$, finding the value which gave the minimum exposure time
needed to detect fading at the 2-$\sigma$ and 3-$\sigma$ levels. These are shown in Fig.~\ref{fig:fadetest}; where we plot
the initial exposure ($E_1$) on the $x$-axis, and take $R_1$ as the count rate at which the source detection algorithm is 50\%\ complete
in that exposure time. If $R_2$ was below the detection threshold in the second observation, we found the exposure time 
that would be needed to give a 3-$\sigma$ upper limit on the count-rate that was below the 1-$\sigma$ lower bound
on $R_1$. We did this for two cases: where the $R_2$ was just below the detection threshold (i.e.\ the source contributed 5 counts
to XRT dataset), and where the source had completely vanished (it contributed no photons); these are plotted
as the dotted lines in Fig.~\ref{fig:fadetest}.

From this we see that a typical repeat observation will need to be \emph{at least} a factor of two longer than the initial exposure;
unless prior knowledge of the source type means that we expect the source to have faded away completely by the time of the second observation.
However, this does not mean that the total observing time needs to be doubled. The goal of the follow-up observations is to determine whether any of the
uncatalogued sources detected in the first observation have faded, not to search for new sources. Therefore, only those fields containing 
uncatalogued sources need to be re-observed. We suggest that a modest investment of observing time spent on observations of the
newly-discovered X-ray sources would significantly increase the likelihood of identifying an X-ray counterpart to a non-EM trigger using \swift\ 
-- provided that the source was detected in the initial observation.

We can determine with more confidence the exposure time needed to 
determine fading if we know in advance what the source light curve looks 
like. One of the main candidates to provide a gravitational wave + EM 
detection are short GRBs, and thanks to \swift\ we have some idea of 
their X-ray light curve morphology. We took from the short GRBs listed 
in \cite{Rowlinson13} those which triggered \swift\ and for which the 
live XRT GRB catalogue contains light curve fits (i.e.\ at least 2 
light curve bins, excluding upper limits). Then, assuming an initial 
exposure of 500 s and 1000 s, starting 1, 2 and 4 hours after the GRB 
trigger, we determined how many of those GRBs would be detected by 
\swift, and at what brightness. We then determined how long a second 
observation would have to be in order to confirm that the source had 
faded at the 3-$\sigma$ level, as a function of the delay between the 
first and second observations. We considered a source to have faded 
either if it was detectable, and had a measured count-rate inconsistent 
with the initial rate at the 3-$\sigma$ level, or if it was undetected 
with a 3-$\sigma$ upper limit inconsistent with that rate. The results 
are shown in the top panels of Fig.~\ref{fig:grbfade}, in which we 
show the exposure time needed to detect fading in 50\%\ and 90\%\ of the 
bursts as a function of the time between observations. Note that ``50\%\ 
of the bursts'' means 50\%\ of the short GRBs that would have been 
detected in the initial observation. The shorter that observation is, or 
the longer after the trigger it begins, the fewer the bursts detected in 
the initial observation. Therefore (for example) the black curve in the 
top-left panel of Fig.~\ref{fig:grbfade} was compiled from 11 GRBs, 
whereas the red curve in the same panel was compiled from 9 
GRBs; the sample from \cite{Rowlinson13} contains 43 bursts\footnote{Since we can't measure fading in objects we don't detect, 
and we are interested in how to detect fading, not how to detect the GRB 
in the first place, this approach is the most informative.}. This explains why 
the necessary exposure is surprisingly short at some times: only the brightest bursts
are detectable, but for these bright bursts, it is relatively simple to identify fading.
For some GRBs, we were not able to confirm fading with 3-$\sigma$ confidence  if 
the delay between observations was too short, with the second observataion's exposure extending to a maximum of $10^5$ 
s; for the purposes of plotting, we set the necessary exposure time in 
these objects to be $10^6$ s (i.e.\ off the scale). For shorter delay 
times it is sometimes not possible to determine fading in even 50\%\ of 
cases; the only solution is to wait longer before performing the second 
observation. Note also that for simplicity we assumed that the second 
observation was a continuous one, i.e.\ we ignored the fact that \swift\ 
can observe a given target for a maximum of 2.7 ks every \til5.7 ks 
orbit.

As noted earlier (Section~\ref{sec:discBright}), the mean redshift of \swift\ short GRBs is \til0.7,
whereas for aLIGO the maximum distance to a short GRB is expected to be 200 Mpc ($z\til0.045$). Therefore we
repeated the above calculations, first shifting the GRB to be at 200 Mpc. We used the redshifts from \cite{Rowlinson13}, i.e.\
reverting to the mean ($z=0.72$) where it is not known. The results of this are shown in the bottom panels of Fig.~\ref{fig:grbfade}.
Not surprisingly, fading is much easier to detect when the GRB is nearby; indeed, if we can detect a GRB at 200 Mpc within 4 hours
of the trigger, then we should be able to confirm fading at most 8 hours later.


\section{Conclusions}
\label{sec:conclusion}

We have used the \swift-X-ray telescope to observe  20 IceCube neutrino doublet triggers,
covering the IceCube error circle in 7 tiled observations. We find 109 X-ray sources, only 16 of which had
been detected in X-rays before these observations were taken. However, none of the uncatalogued sources are bright enough
to be distinguished from the serendipitous sources expected in a 7-tile XRT observation. Given the
behaviour of GRB afterglows observed by the \swift-XRT, this lack of bright counterparts allows us to rule out with $>99\%$ confidence that
the neutrino triggers arise exclusively from GRBs. In only one case could we identify signs
of fading in these single observations, and follow-up observations showed that this was not
a transient object.

Considering the wider question of using \swift-XRT to detect the counterpart to non-EM triggers, such as PeV neutrinos or
Gravitational Wave triggers from the Advanced-LIGO facility, we have shown that a deeper all-sky X-ray catalogue 
than the RASS, such as that which will be created by \emph{eRosita}, will make it easier
to identify a counterpart simply from its flux in the initial observation. However, a better 
and more immediately available approach will be to probe source variability, in particular, searching for fading.
Re-observing every uncatalogued X-ray source detected by XRT in the initial follow-up observations, with at least twice the initial exposure time, 
should enable us to identify any source which has faded at the 3-$\sigma$ level.

For a short GRB at a distance of 200 Mpc -- as aLIGO expects to discover -- with its
jet pointed towards Earth, an initial observation within a few hours of the trigger, followed
by a second observation of 2-ks, 8 hours after the first, should be sufficient
to confirm fading in almost all cases (if the GRB was detected in the initial observation),
provided the afterglows of such GRBs are of the same luminosity
and morphology distributions as the sample of short GRBs detected to date by \swift.

While we have demonstrated techniques to maximise the potential for multi-messenger
astronomy with \swift, a sensitive, all-sky (or at least large field-of-view)
X-ray imager, ideally with some form of gamma-ray detector to rapidly distinguish GRBs from other X-ray transients,
would be the ideal facility with which to locate the high-energy counterparts to non-EM-detected transients. 

\section*{Acknowledgements}

This work made use of data supplied by the UK Swift Science Data Centre at the University
of Leicester. PAE and JPO acknowledge UK Space Agency support. JMG gratefully acknowledges
the support from NASA under award NNH13CH61C. We thank the anonymous referee for their helpful
and constructive feedback on the manuscript.

\bibliographystyle{mn2e}
\bibliography{evans_swift_ic}

\label{lastpage}
\end{document}